\documentclass[twocolumn,aps,10pt,superscriptaddress,showkeys,noeprint]{revtex4-2}
\pdfoutput=1
\usepackage{style}

\begin{document}

\title{Capturing the diversity of multilingual societies}

\author{Thomas Louf}
\email{thomaslouf@ifisc.uib-csic.es}
\author{David S\'{a}nchez}
\author{José J. Ramasco}
\affiliation{Institute for Cross-Disciplinary Physics and Complex Systems IFISC (UIB-CSIC), 07122 Palma de Mallorca, Spain}

\keywords{computational sociolinguistics $|$ language dynamics $|$ bilingualism $|$ agent-based models $|$ Twitter data} 

\date{\today}

\begin{abstract}
Cultural diversity encoded within languages of the world is at risk, as many languages have become endangered in the last decades in a context of growing globalization. To preserve this diversity, it is first necessary to understand what drives language extinction, and which mechanisms might enable coexistence. Here, we study language shift mechanisms using theoretical and data-driven perspectives. A large-scale empirical analysis of multilingual societies using Twitter and census data yields a wide diversity of spatial patterns of language coexistence. It ranges from a mixing of language speakers to segregation with multilinguals on the boundaries of disjoint linguistic domains. To understand how these different states can emerge and, especially, become stable, we propose a model in which language coexistence is reached when learning the other language is facilitated and when bilinguals favor the use of the endangered language. Simulations carried out in a metapopulation framework highlight the importance of spatial interactions arising from people mobility to explain the stability of a mixed state or the presence of a boundary between two linguistic regions. Further, we find that the history of languages is critical to understand their present state.
\end{abstract}

\maketitle

Language, as the basis for communication, is at the heart of the functioning of human societies. It has thus long been an important subject of research, as scientists sought to understand its interactions with society, the internal evolution of a language's aspects with time or how multiple languages interact with one another. The research presented here is concerned with the latter, which emerged a few decades ago as a hot topic when linguists realized that the world may be facing a mass extinction of languages \cite{Krauss1992,Grenoble1998,Crystal2000}. It has been pointed out that the estimated \SI{6000}{} languages of the world convey a cultural wealth, the loss of which would be irreversible. Hence the need to understand what drives individuals to shift from one language to another.

Modeling language shift has been the subject of much research in the last decades \cite{Castellano2009,Boissonneault2021}, which employed various approaches such as the formulation of evolution equations based on ecological models \cite{Mira2005, Pinasco2006, Kandler2008, Sole2010,Heinsalu2014}, of reaction-diffusion equations \cite{Kandler2009,Patriarca2009,Isern2014,Prochazka2017}, or approaches within the framework of agent-based modeling \cite{Castello2006, Minett2008,Castello2013,Prochazka2017}. While global evolution equations determine how the proportions of each language group will evolve in a system, agent-based models (ABMs) describe the shifting mechanisms on an individual level, as they provide probabilities to switch to another language group. These transition probabilities depend on the linguistic environment of the individual, environment which may be defined in many ways. Different networks of interactions can be introduced, ranging from the simplest (fully-connected networks) to more realistic but less tractable ones (like a real-world social network). The former lend themselves easily to mathematical analysis as they can be equivalently written in terms of global evolution equations for large population sizes. As a result, models based on global evolution equations are a subset of the more general, agent-based ones. Moreover, ABMs allow to assess the impact of the social structure on the dynamics. This social structure is closely related to space, but in a non-trivial way, and as there is no model that can claim to be the universal solution to build spatial interaction networks \cite{Barbosa2018}, being able to plug in any kind of interaction network is an interesting feature of ABMs. It is for all these reasons that the focus of this article will be on ABMs.
The first notable model to mention is the Abrams-Strogatz model \cite{Abrams2003}. It was the first to attract considerable attention as the authors were able to fit their model to the historical data of multiple languages threatened by extinction, and subsequently predicted their death. The model is very simple as it considers only the monolingual states A and B. The basic principle behind this model is that the more speakers of A, and the more prestigious A is in society, the more B speakers will want to switch to A, and inversely.

However, the existence of around \SI{6000}{} spoken languages in 200 nations implies that multilingualism is a pervasive phenomenon worldwide. In almost every country, the presence of more than one language naturally leads to speech communities of different sizes. A common situation is that many individuals belonging to these communities use two or more languages independently of the official status and the educational prevalence of those languages. The extent and role of bilingualism is hence a difficult subject. Multiple modeling attempts have been made in that direction \cite{Castello2006,Patriarca2009,Patriarca2012,Vazquez2010}. In these models, agents can be in a third state AB through which they have to pass to switch from being monolingual in a language to another. Apart from \cite{Prochazka2017} which relied on census data, none of the aforementioned models have been iterated over real-world spatial distributions of speakers, as they were rather implemented in fully-connected populations or in toy models, like lattices or random networks. This is a shortcoming we will address here.

Indeed, speech communities are distributed in regions which are heterogeneous and even discontinuous when their boundaries cannot be arranged into a single closed curve. This spatial component cannot be neglected in the study of language dynamics, as the sociolinguistic environment in which individuals interact is of paramount importance for the dynamics. That is why this work also seeks to obtain and analyze the spatial distribution of languages in order to evaluate the models. %Indeed, they have seldom been tested using such data. 
But despite the ubiquity of language, data on language use have historically been hard to come by. Linguists have mainly relied on data from censuses or surveys which have a limited scope, especially in terms of spatial resolution and sample size. Thus, \cite{Nguyen2016} argued for large-scale data-driven approaches to complement existing sociolinguists' works, in a complementary framework of ``computational sociolinguistics''. In addition to new tools for speech and text analysis, technological advancements have brought with them the ability to collect unprecedented amounts of data from online communications.

In this work, we combine a large-scale empirical study of the spatial distribution of languages with agent-based modeling. In Sec. II, we show empirically that multilingual societies are characterized by different spatial patterns in the populations of monolinguals and bilinguals, encompassing fully mixed states and segregated distributions with a clear linguistic boundary. As the existing ABMs are not able to explain the range of spatial mixing observed, we introduce in Sec. III a model able to capture the diversity seen in the data. The model also shows how the behavior of bilinguals and the ease of learning a language have their importance for the coexistence of languages. Finally, Sec. IV contains our conclusions.

\section{A diversity of multilingual societies?}
As said above, multilingual societies are numerous and thus susceptible to display distinct features. These differences, however, need to be observed and, ideally, quantified, to truly describe the diversity of these societies. Given the very few regions and countries where censuses gather data on language use at a fine enough spatial scale, we choose here to turn to Twitter as an alternative data source. Nonetheless, our analysis can equally be applied to data from surveys and census where available, as shown in the Supplementary Information (SI) Sec. II and Figs. S13 and S14 for Quebec \cite{supp}.

\subsection{Twitter data analysis}
\begin{figure*}[t!]
    \centering
    \includegraphics{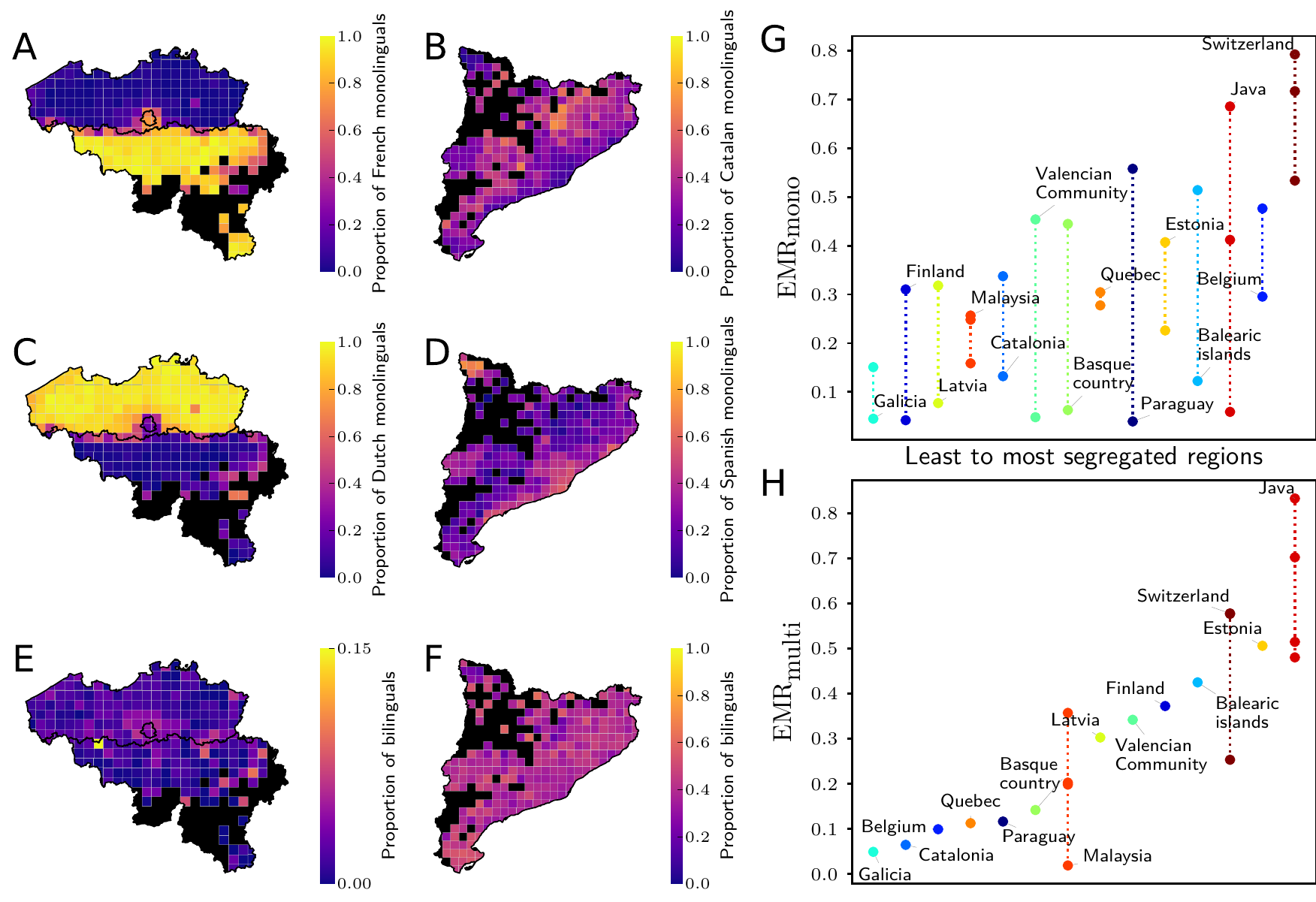}
    \caption{Visualization of the diversity of multilingual societies. For each cell of $10 \times 10 \, \si{\kilo \meter \squared}$, the proportions $p_{L,i}$ of monolinguals in (\textit{A}) French, (\textit{B}) Catalan, (\textit{C}) Dutch and (\textit{D}) Spanish in Belgium (left) and Catalonia (right) are shown. The maps (\textit{E}) and (\textit{F}) show the proportion of bilinguals (note the different scale needed in (\textit{E})). In the case of Belgium, the border between Flanders (North) and Wallonia (South) is drawn, and the Brussels Region too. In black are cells in which fewer than 10 Twitter users speaking a local language were found to reside, consequently discarded for the insufficient statistics. A clear separation of language groups is visible in Belgium following the linguistic regions, displaying mixing mainly around the border and in Brussels, while mixing in Catalonia is much more widespread, with a slight difference between the countryside and the large cities of the coast (East). (\textit{G})-(\textit{H}) Earth Mover's Ratios of respectively the monolingual and multilingual groups of multilingual regions of interest, ranked left to right by increasing average of the $y$-axis values. In (\textit{H}), the point for trilinguals in Switzerland is not displayed because its value was deemed unreliable (for more details see SI Sec. IF \cite{supp}). A rich diversity of mixing patterns is shown, beyond the two paradigmatic cases of Catalonia and Belgium.}
    \label{fig:BE_CAT_maps_EMD_scatter}
\end{figure*}

Twitter is a social networking and micro-blogging service used worldwide by hundreds of millions of users, who post short messages, called tweets, which can be geo-located. It has thus good potential as a data source to extract spatial distributions of language use, as shown in \cite{Mocanu2013,Pavalanathan2015,Goncalves2014,Huang2016,Goncalves2018,Dunn2020}.
Here, we are not so much interested in language distributions fitting perfectly what exists in the offline world, but rather in the kind of distributions we may encounter.
Despite all the biases introduced by the differences of usage of Twitter across the population, it could hence still provide valuable insights for regions in which close to no other data are available.
Then to obtain spatial distributions of languages, we selected 16 countries and regions in which there was potential to gather sufficient statistics for multilingual communities (see the list in the Table S1 of the SI \cite{supp}), and analyzed geo-located tweets sent from them from early 2015 to the end of 2019. A regular grid was laid over each area of interest, dividing them in square cells (see for instance the grids laid over Belgium and Catalonia in Fig. \ref{fig:BE_CAT_maps_EMD_scatter}). The cell size has to be adapted for each studied region, as explained in Sec. I C of the SI \cite{supp}. We have checked the effect of modifying the cell size and made sure that our results are robust (see Supplementary Figs. S10, S11 and S12 \cite{supp}). 
The language of the messages is detected using Chromium's Compact Language Detector (CLD) \cite{Solomon2014} that provides the most likely language of a text from the messages along with a confidence (see Sec. IC of the SI \cite{supp} for details). After thoroughly cleaning and analyzing the collected tweets, we obtained a sample of local Twitter users to which a cell of residence and a set of languages were attributed. Information about data access and code availability can be found in the Appendix.

\subsection{Metrics}
Before introducing any metric, we specify our definition of language groups. 
First, we focus only on the languages considered local. For instance, the use of English is widespread on Twitter, but we do not register those tweets unless English is one of the local languages (e.g., in Canada or Malaysia). A user is classified as a speaker of a language if at least $10\%$ or 5 of their tweets are detected in that language. One individual can thus be naturally in a monolingual or in a multilingual group if they fulfill the condition in more than one language. The groups defined here are mutually exclusive: each user must be in one of the monolingual and multilingual groups that are possible to form with the given set of local languages. For the purposes of our work, we consider language as a social phenomenon. Thus, we do not take into account the individual proficiency, which is indeed interesting in other fields of study \cite{Baker2011}, but instead observe the language production of a speech community defined inside every cell, based on their use of one or more languages. Thereafter, we will talk of $L$-speakers instead of ``individuals who belong to the $L$-group'' for simplicity.

Starting from the counts $N_{L,i}$ of $L$-speakers residing in cell $i$ obtained from the data, we wish to gain insights on the spatial distributions of language use. To do so we need to define a few basic metrics:
\begin{itemize}
       \item concentration in cell $i$ of $L$-speakers:
       \begin{equation}
       \label{eq:def_conc}
              c_{L,i} = \frac{N_{L,i}}{N_L},
       \end{equation}
       \item proportion of $L$-speakers in $i$'s population:
       \begin{equation}
       \label{eq:def_prop}
              p_{L,i} = \frac{N_{L,i}}{N_i},
       \end{equation}
\end{itemize}
where $N_L$ are all the users classified as $L$-speakers in the country or region considered, and $N_i$ is the population of Twitter users residing in cell $i$ speaking any of the local languages.
As in \cite{Mocanu2013}, we can define the polarization of a language A for every cell $i$ in a bilingual system with languages A and B as 
\begin{equation}
\label{eq:def_polar}
    \theta_{A, i} = \frac{1}{2} (1 + p_{A,i} - p_{B,i}). 
\end{equation}
The polarization vanishes when there are only B monolinguals, takes the neutral value of $0.5$ when there are as many A-speakers as B-speakers, and goes to $1$ when there are only A monolinguals. We will use this metric in bilingual regions as an indication of the mixing at the cell level. 

Building further upon proportions and concentrations, we want to be able to measure the spatial mixing of language groups, or inversely, their spatial segregation. We define segregation as the difference in how individuals of a given group are spatially distributed compared to the whole population. Segregation is thus conceptualized as the departure from a baseline, the unsegregated scenario, in which regardless of the group an individual belongs to, they would be distributed according to the whole population's distribution. Explicitly, the concentrations corresponding to this baseline, or null model, are $c_i = N_i / N$. To quantify language mixing, we would then like to measure a distance between the spatial distribution of a given language group and that of the whole population.

To this end, at a full country or region scale, we define the so-called Earth Mover's Distance (EMD). This metric allows us to quantify the discrepancy between two distributions embedded in a metric space of any number of dimensions. It has mainly been used within the field of computer vision \cite{Rubner1998}, and it was shown to be a proper distance (in the metric sense) between probability distributions \cite{Levina2001}. Here, we consider the distributions defined by the signatures $P = \{ (i, c_i) \}$ and $Q_L = \{ (i, c_{L,i}) \}$. We then define $\text{EMD}_L$ as 
\begin{equation}
\label{eq:emd_def}
    \text{EMD}_L \equiv \text{EMD}(P, Q_L) = \sum_{i,j} \hat{f}_{ij} d_{ij},
\end{equation}
with $d_{ij}$ the distances between cells $i$ and $j$, and $\hat{f}_{ij}$ the optimal flows to reshape $P$ into $Q_L$, obtained by minimizing $\sum_{i,j} f_{ij} d_{ij}$ under the following constraints:
\begin{equation}
    \left\{
    \begin{aligned}
        f_{ij} & \geq 0, \forall \, i,j \\
        \sum_j f_{ij} & = c_{L,i}, \forall \, i \\
        \sum_i f_{ij} & = c_j, \forall \, j \\
        \sum_{i} \sum_j f_{ij} & = \sum_i c_{L,i} = \sum_j c_j = 1,
    \end{aligned}
    \right.
\end{equation}
where $c_i$ and $c_{L,i}$ are the concentrations of the population and $L$-speakers in every cell $i$, as defined above. $\text{EMD}_L$ quantifies thus the distance between the concentration distributions of $L$-speakers and of the whole population, as needed. The computation of the EMD was implemented with \cite{Flamary2021}, which uses the method of \cite{Bonneel2011}. However, in its raw form, it is dependent on the spatial scale of the system considered. Hence the need for a normalization factor $k_\text{EMD}$ in order to enable comparisons between regions of different sizes. The first, obvious choice for $k_\text{EMD}$ would be the maximum distance between two cells of the region. However, such a choice would neglect the disparities of population density existing between different regions. The factor would be very high in Quebec, for instance, since the geographical scales are large even though its northern part is scarcely populated. This is why we choose instead the average distance between individuals:
\begin{equation}
    k_\text{EMD} = \frac{\sum_i \sum_j N_i N_j d_{ij} }{\left( \sum_k N_k \right)^2}.
\end{equation}
Our final metric is then the normalized version of the EMD, the EMR (Earth Mover's Ratio), defined as:
\begin{equation}
\text{EMR}_L = \frac{\text{EMD}_L}{k_\text{EMD}} .
\end{equation}
The EMR is a global parameter. The higher it is, the more segregated a linguistic community. On the contrary, if the EMR is close to zero this community is distributed according to the total population and the mixing is complete. As shown in the SI Fig. S13 and Sec. S14 \cite{supp}, the EMR is cell size invariant and, quite generally, a reliable metric when a careful statistical analysis is made.

\subsection{Empirical results}
We propose a first visualization of the collected data in Fig. \ref{fig:BE_CAT_maps_EMD_scatter}\textit{A}-\textit{D}, where the proportions of monolinguals in Dutch and French, Catalan and Spanish, are displayed for Belgium and Catalonia, respectively. The cell size is here of $10 \times 10 \, \si{\kilo \meter \squared}$ (see Supplementary Figs. S10 and S11 \cite{supp} for equivalent maps with cells of $5 \times 5 \, \si{\kilo \meter \squared}$ and $15 \times 15 \, \si{\kilo \meter \squared}$). The maps already show two configurations that frequently appear across the world in multilingual societies: either a marked boundary between mostly monolingual domains (Belgium) or high mixing in every cell with local coexistence (Catalonia). The population of bilingual users concentrate in the border in the first case (especially in the region around Brussels and in the southern border with Luxembourg), and it is widespread in the second (Fig. \ref{fig:BE_CAT_maps_EMD_scatter}\textit{E}-\textit{F}).
Results for the other multilingual regions listed in the SI Table S1 are shown in the SI Figs. S1-S14 \cite{supp}. These findings are summarized in Fig. \ref{fig:BE_CAT_maps_EMD_scatter}\textit{G}-\textit{H}, which presents the ranges of values reached by the EMR of respectively the monolingual and multilingual groups in 14 of our 16 regions of interest. We filtered out regions where we deemed not sufficient the statistics gathered from Twitter (see the SI Table S2 for all measured metrics and cell sizes used \cite{supp}). A wide diversity of situations can be observed.
Multilingual societies may have rather balanced monolingual groups separated by a clear-cut border, which have thus high but quite similar EMR values, like in Belgium and Switzerland. One can also see unbalanced situations where one language is majoritarian, and has thus a much lower EMR than the monolinguals and multilinguals of other smaller, isolated languages. This is for example the case on the island of Java, where Indonesian is widespread, and Javanese and Sundanese are more localized. Multilinguals may also be mixing well in the whole population, like the bilinguals in Galicia and Catalonia. These groups can thus be of completely different natures from one region to another, from sustaining a minority language while being spatially mixed or isolated, to standing at the border between monolingual communities.

The metrics introduced to evaluate the spatial mixing of languages can be calculated using similar data taken from other sources. Although data on language use on a fine enough spatial scale are difficult to find, it can, for instance, be obtained for Quebec from the Canadian census of 2016. Maps equivalent to the ones of Fig. \ref{fig:BE_CAT_maps_EMD_scatter} are shown using both data from the census and from Twitter for Quebec in the SI Figs. S13 and S14 \cite{supp}. Similar mixing patterns can be observed from both data sources.

\section{Models capturing diversity} 
As language use in a society only sees significant changes on a time scale of generations \cite{Labov1973}, the maps obtained from Twitter are only snapshots of the situation around the years 2015 to 2019 (synchronic viewpoint). We do not have access to data providing the longitudinal evolution (diachronic framework), but the models at hand do describe the dynamics of the system. Since some of the multilingual societies we study have had the same kind of spatial pattern of language coexistence for generations (Belgium with a separation and Catalonia with mixing), it is natural to ask whether these states are stable solutions of a model describing language competition. We will check, in the first place, if the existing models meet the basic requirement of reaching the observed stable states. Crucially, if they do not fulfill it, the underlying mechanisms of language shift are not therein fully captured, missing a significant element that could be key to language preservation.

\subsection{Previous models}
\begin{figure}
\centering
    \includegraphics[width=\linewidth]{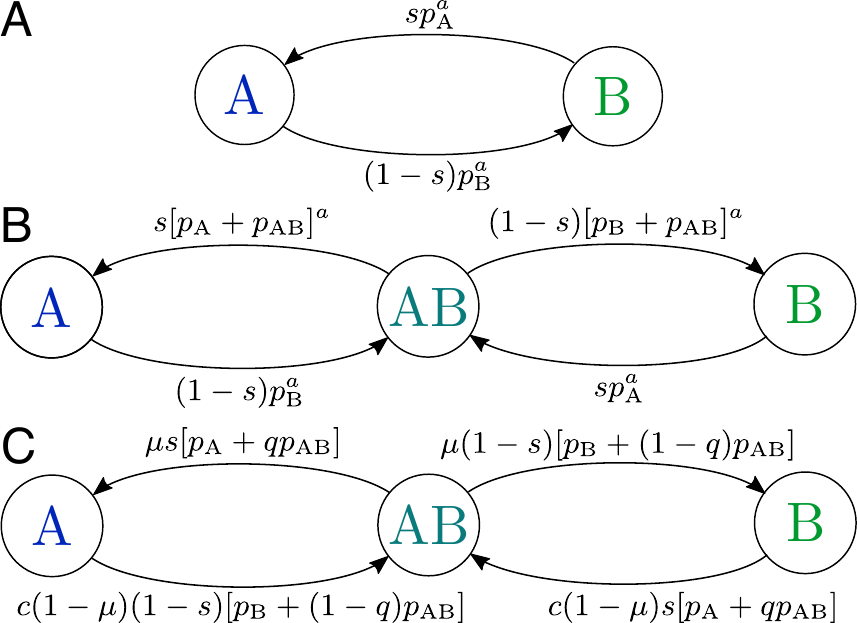}
    \caption{Diagrams of the models presented in the text, showing the transition probabilities from one state to another. (\textit{A}) Abrams-Strogatz model from \cite{Abrams2003}. (\textit{B}) Bilinguals model from \cite{Castello2006}. (\textit{C}) Our model of bilinguals including both their preference and the ease to learn the other language (see equation \eqref{eq:bipref_model}).}
    \label{fig:models_diagram}
\end{figure}

The individuals in a population can be in states representing their use of one or several languages. Under this framework, the dynamics are governed by the permitted transitions between states and their corresponding probabilities of occurring. Fig. \ref{fig:models_diagram} displays the states: monolingual in A and B, and bilingual AB, with the associated transition probabilities in two previous models and in our proposal. We denote $p_{\text{A}}$ and $p_{\text{B}}$ the proportions of monolinguals in A and B, respectively, and $p_{\text{AB}}$ the proportion of bilinguals. Within a mean-field approximation, and all the population being mixed, all equations can be written in terms of the proportions, which satisfy the equality $p_{\text{A}} + p_{\text{B}} + p_{\text{AB}} = 1$. Within this notation, a state of coexistence is a state in which the two languages remain spoken, which corresponds to either $p_{\text{AB}} > 0$, or $p_{\text{A}} > 0$ and $p_{\text{B}} > 0$. Extinction of A (B), for instance, corresponds to $p_{\text{A}} = p_{\text{AB}} = 0$ ($p_{\text{B}} = p_{\text{AB}} = 0$).
% mixing implies coexistence, while coexistence does not imply mixing (in spatial distribution: linguistic communities may mix in the same cells, or be completely separated, but there is still global coexistence). -> mixing = local coexistence

The first model to mention is the one introduced in Ref. \cite{Abrams2003} by Abrams and Strogatz (Fig. \ref{fig:models_diagram}A). The model only contains monolinguals, who can change their languages with a probability that depends on the proportion of speakers of the other language to an exponent $a$ (called volatility), which controls if the dependence on the proportion of the other language group is linear ($a = 1$), sub-linear ($a<1$) or super-linear ($a>1$). Besides, they also include a parameter $s$ between zero and one, which stands for the prestige of the language A. If $s$ is close to one, all the individuals will forget B and start to speak A alone. Set in a single population and in mean field, this model was shown to fit historical data of the decline of minority languages in \cite{Abrams2003}. It was thoroughly analyzed in \cite{Vazquez2010}, where it was first shown that its stable state is extinction of one language for $a \ge 1$, and coexistence for $a < 1$, independently of the prestige. In complex contact networks, the coexistence region in the $(s, a)$ space shrinks, as not all values of prestige enable coexistence for $a < 1$. It is important to note that the linear version of the model does not predict coexistence.    

Later, an extended model with bilinguals was proposed by Castell\'o et al. \cite{Castello2006} (see Fig. \ref{fig:models_diagram}\textit{B}). The transitions to lose a language are there related to the proportion of bilinguals besides the monolinguals of the other side. The idea is that since A can be spoken to both A and AB individuals, the utility to retain B decreases with an increasing proportion of these two types of individuals. An analysis of the stable states of this bilinguals model performed in Ref. \cite{Vazquez2010} shows that the coexistence only occurs if $a < 1$ and that the area of parameters allowing it is reduced compared to the Abrams-Strogatz model. Again, the linear ($a = 1$) version of the model does not allow for language coexistence.

Several concerns may be raised about these models. The first one is that for languages with equal prestige ($s=1/2$) and with equal social pressure (same proportion terms), learning and forgetting a language is equiprobable, while they result from two completely different processes. People may inherit a language from their parents, use it for endogenous communication, and they could be driven to learn a new one for work or education purposes, which corresponds to exogenous communication. This is a typical diglossic situation \cite{Ferguson1959} with a linguistic functional specialization. A difference in prestige favors this process, but losing a language, especially in the presence of cultural attachment, can be more difficult. In the case of bilingualism, once someone masters a new language to a bilingual level they will not forget their first. Besides, it seems reasonable to assume that most of the time, a language is lost when it is not passed from one generation to the next \cite{Crystal2000, Portes1998}. 
A second concern we raise here is that both models only find stable coexistence in a nonlinear configuration, when $a < 1$. These values of $a$ imply easier transitions overall, and thus that coexistence is favored when speakers are more loosely attached to their spoken languages. This nonlinearity is hence hard to explain from a practical point of view and it has the effect of making the transitions less dependent on the actual proportions of speakers. Thirdly, it is important to note that the bilingual model of Fig. \ref{fig:models_diagram}\textit{B} is not able to produce a stable solution in which the bilinguals coexist with monolinguals of a single language.

\subsection{Our model}
Our proposal stems from the realization of this last point: there are several bilingual societies where the monolinguals of one language, e.g., B, are virtually extinct (e.g., Catalonia, Quebec or the Basque Country). However, the bilinguals continue to use B and keep it alive for decades if not centuries due to cultural attachment. This ``reservoir effect'' must be incorporated in models of language shift. The other ingredient that we will include concerns demographics, in relation with the first concern raised above: language loss mostly occurs between generations. For this, we get inspiration from the work of Ref. \cite{Minett2008} that sets a rather generic framework for models differentiating horizontal and vertical transmission.

We thus first distinguish generational, or vertical, transmission, which corresponds to the death of a speaker replaced by their offspring. If the speaker was monolingual, their single language is transmitted. If they were bilingual, one of their two languages might get lost in the process of transmission. This loss occurs according to the following transition probability:
\begin{equation}
    P (\text{AB} \rightarrow \text{X}) = \mu \, s_\text{X} \, \left[ p_{\text{X}} + q_\text{X} \, p_{\text{AB}} \right],
\end{equation}
where, as in the other models, $s_\text{X}$ refers to the prestige of language X, which can be either A or B. The other parameters are $\mu \in [0, 1]$, that is the fixed probability for an agent to die at each step; and, $q_\text{X} \in [0, 1]$ that reflects the preference of bilinguals to speak X. So bilingual speakers may be more inclined to transmit only language X when it is more prestigious, preferred by other bilinguals, and more spoken around them.

The second kind of transition is horizontal, it is related to the learning of a new language by a monolingual in the course of their lives. This transition occurs according to the following transition probability:
\begin{equation}
    P (\text{X} \rightarrow \text{AB}) = c \, (1 - \mu) \, s_\text{Y} \, \left[ p_\text{Y} + q_\text{Y} \,  p_{\text{AB}} \right],
\end{equation}
where Y is the language other than X, and, critically, $c \in [0, 1]$ is a factor adjusting the learning rate. The time scales of the learning process and of a generational change are completely different, hence the need to adjust $(1-\mu)$ by this factor $c$ here. It depends on the similarity between the two languages and on the implemented teaching policies. For the sake of simplicity and to avoid the inclusion of more parameters, we assume that the process is symmetric between learning A when B is spoken and vice versa. This is not necessarily true in all cases, but it can easily be solved by splitting $c$ in more parameters for each transition. To translate this expression of the transition probability into words, a monolingual in X will be more willing to learn Y as it is easier to learn, more prestigious, preferred by bilinguals, and more spoken around them.

We define $s$ and $q$ as symmetric around $1/2$, and thus define $s = s_\text{A} = 1 - s_\text{B}$ and $q = q_\text{A} = 1 - q_\text{B}$. The transitions in our model are illustrated in Fig. \ref{fig:models_diagram}\textit{C} and we explicit here below the transition probabilities that define it:
\begin{equation}
\label{eq:bipref_model}
\left\{
\begin{aligned}
    P (\text{A} \rightarrow \text{AB}) &= c \, (1 - \mu) \, (1-s) \, \left[ p_{\text{B}} + (1-q) \,  p_{\text{AB}} \right] \\
    P (\text{B} \rightarrow \text{AB}) &= c \, (1 - \mu) \, s \, \left[ p_{\text{A}} + q \, p_{\text{AB}} \right] \\
    P (\text{AB} \rightarrow \text{A}) &= \mu \, s \, \left[ p_{\text{A}} + q \, p_{\text{AB}} \right] \\
    P (\text{AB} \rightarrow \text{B}) &= \mu \, (1-s) \, \left[ p_{\text{B}} + (1-q) \, p_{\text{AB}} \right]
\end{aligned}
\right. .
\end{equation}
An important aspect of the model is that the use of a language by bilinguals contributes potentially unequally to the sizes of each language community. The neutral case occurs when $q = 1/2$ and bilinguals on average contribute equally to both groups. It is however natural that even if bilinguals are fluent in both languages, individually they may have a certain preference for one of them and their language use is not necessarily balanced \cite{Romaine2012}. Even if one of the two languages is in a minority or suffers from a lack of prestige, appropriate values of $q$ may maintain it alive. The most extreme example occurs when the monolinguals of B, for example, are extinct ($p_{\text{B}} = 0$). Still, the use of B by the bilinguals keeps attracting monolinguals of the group A proportionally to $(1-q)\, p_{\text{AB}}$.

Finally, we chose not to include nonlinearities in the model ($a = 1$), as it turned out not to be necessary to capture the diversity we observed, and it would only add unnecessary complexity.

\subsection{A single population}

\begin{figure*}
    \centering
    \includegraphics[width=\textwidth]{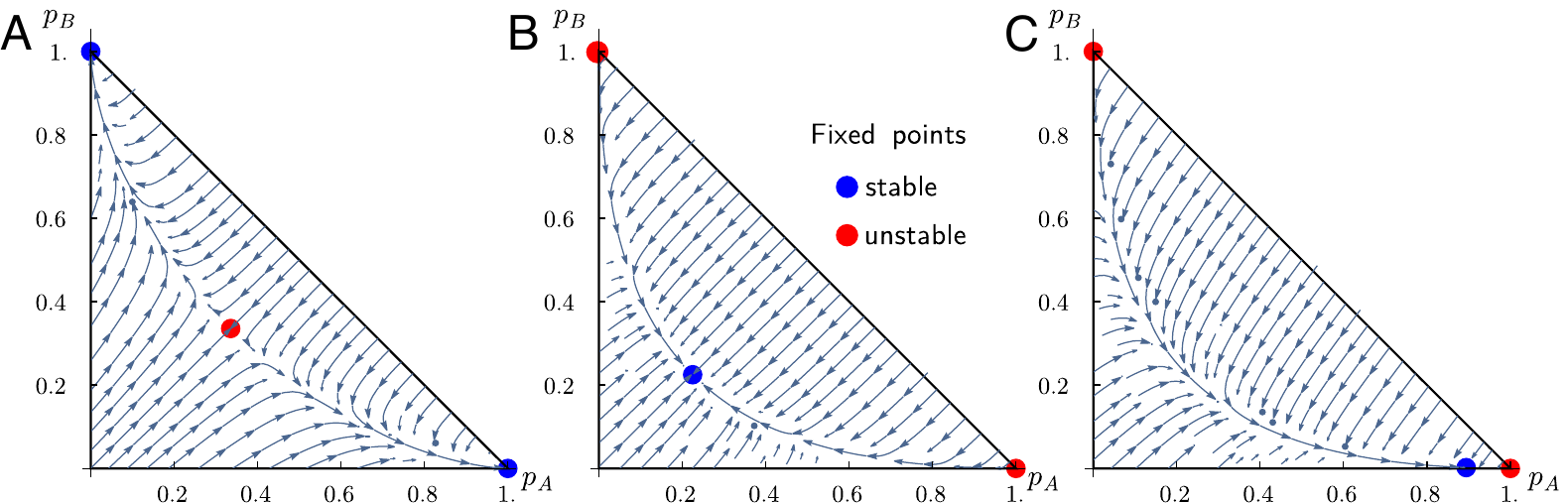}
    \caption{Flow diagrams for the dynamics of two languages according to our model described in equation \eqref{eq:bipref_model} set in a well-mixed population. $p_{\text{A}}$ and $p_{\text{B}}$ denote the proportions of monolinguals in A and B, respectively, and the proportion of bilinguals $p_{\text{AB}}$ is such that $p_{\text{A}} + p_{\text{B}} + p_{\text{AB}} = 1$. The mortality rate is fixed at $\mu = 0.02$. (\textit{A}) For $s = q = 1/2$ and $c=0.02$, the stable outcome is extinction of one of the two languages. (\textit{B}) For $s = q = 1/2$ and $c = 0.05$, the higher learning rate leads to a solution featuring stable coexistence. (\textit{C}) For $s = 0.57$, $q = 0.45$ and $c = 0.05$, despite the lower prestige, B survives in a small community of bilinguals as it is the preferred language among them.}
    \label{fig:stream_plots}
\end{figure*}

We first analyze the model in the simplest setting of a single well-mixed population
to determine the typology of possible solutions. Given the normalization condition $p_\text{A}+p_\text{B}+p_\text{AB} = 1$, the system dynamics can be described by a set of two coupled equations, let us say, for $p_\text{A}$ and $p_\text{B}$ (see the SI Sec. III \cite{supp}). Fixed points are the solutions for which $\partial p_\text{A}/\partial t = \partial p_\text{B}/\partial t = 0$. The stability of these points is studied by performing a linear perturbation analysis around them, which requires the calculation of the Jacobian of the linearized equations and of its eigenvalues. Points for which all the eigenvalues have strictly negative real parts are stable, while if any eigenvalue's real part is zero or positive the fixed point is unstable. Stream plots in Fig. \ref{fig:stream_plots} show where the model converges to in three characteristic examples, depending on the model parameters. In the first one (Fig. \ref{fig:stream_plots}\textit{A}), the stable (blue) points lie over the axis at values 1 and the system has as only solution the extinction of one of the two languages. In Fig. \ref{fig:stream_plots}\textit{B}, the stable fixed point falls in the middle of the diagram and, therefore, the solution is symmetric coexistence with a majority ($\sim 1/2$) of bilinguals. Finally, in Fig. \ref{fig:stream_plots}\textit{C}, we find a stable fixed point over the $x$-axis that represents the extinction of monolinguals B but coexistence between A-monolinguals and bilinguals. Surprisingly enough, this represents the survival of a less prestigious language within a relatively small bilingual community.
These results show already the flexibility of the model even in a single population.

\begin{figure}[h]
\centering
    \includegraphics[width=7cm]{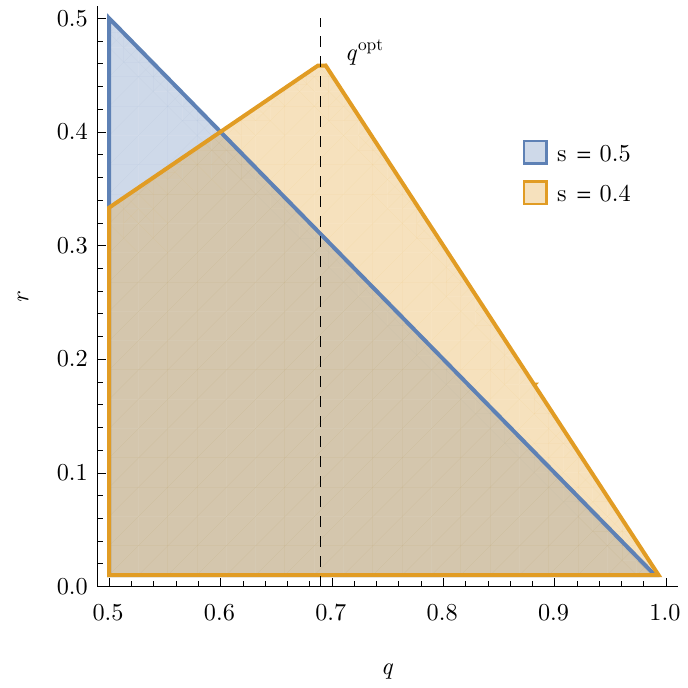}
    \caption{Region of the parameter space where the dynamics of our model in a single population converge to stable coexistence of languages. We show two 2D cuts of the coexistence region in the ($q$, $r$) space for fixed values of $s=0.5, 0.4$, with $r = \mu / (c \, (1-\mu))$. Lower values of $r$ favor coexistence, as well as a neutral prestige and bilingual preference $q$. When $s < 0.5$, coexistence is favored for an optimal value $q^\text{opt} > 1-s$.}
    \label{fig:coex_region}
\end{figure}

We change now the viewpoint from the phase space to the parameter space. In Fig. \ref{fig:coex_region}, we plot the region of parameters where the model converges to stable coexistence. Since $c$ and $\mu$ act over the stability only in a combined form, their contributions can be merged into a new variable $r$ defined as $r = \mu/(c \, (1-\mu))$, which stands for the ratio between the mortality and learning rates. The other two parameters, $s$ and $q$, are considered independently. We observe that the coexistence region expands when $r$ decreases. This means that increasing the ease to learn one language when knowing the other (with a fixed mortality rate) makes coexistence more likely. Additionally, coexistence occurs more frequently when both prestige and bilingual preference are neutral, $s = q = 1/2$, which is expected. When the prestige of language A is lower than that of B, we find that there exists an optimal value of $q$ making possible the coexistence, $q^\text{opt} > 1-s$. 
For $q < q^\text{opt}$, $A$ is more at risk of extinction whereas for $q > q^\text{opt}$, the endangered language is $B$. There is thus a balance between prestige and bilingual preference that enables coexistence. 

This model opens up unique classes of stable solutions: from the extinction of a language to coexistence when prestige is neutral, but also when it favors one of the two languages, and even only through a community of bilinguals. However, these analytic results in a fully-connected population do not suffice, as they do not show if the model is able to reproduce a case such as Belgium, where in the majority of cells there remains almost exclusively one language, except on the boundary between the two large communities. Consequently, we now analyze the model in a metapopulation framework to uncover the effect of including space and check whether this pattern can arise.

\subsection{The model in space}

The idea of introducing a metapopulation framework in order to study interaction dynamics in space has been extensively exploited in ecology \cite{Hanski1998} and epidemiology \cite{Sattenspiel1995,Balcan2010}. In our context, we would need some information to build the extended model. The basic ingredients are a spatial division, the population in each division, the mobility between them and the characteristics of the populations in terms of language groups. Since we are interested in the phase space of the model, it is possible to use a completely abstract setting. However, this would require the generation of reasonable data in terms of population and mobility, while this information is easily accessible from census data in many countries. Since we wish here to study the stability of the present, observed state, to make metapopulations interact with one another we use readily-available commuting data from the census, as commuting is the backbone of everyday mobility. Some further work could include other kinds of mobility, like migrations, in order to investigate long-term time evolutions. We have thus chosen to use census data in Catalonia and Belgium as benchmarks, although it is important to stress that the intention is not to produce accurate predictions. Alternatively, the spatial interactions could be estimated from the population data using a model of human mobility, such as gravity, radiation or distance-kernel-based models \cite{Barbosa2018,Burridge2017,Burridge2021}.

\begin{figure}[h]
\centering
    \includegraphics[width=\linewidth]{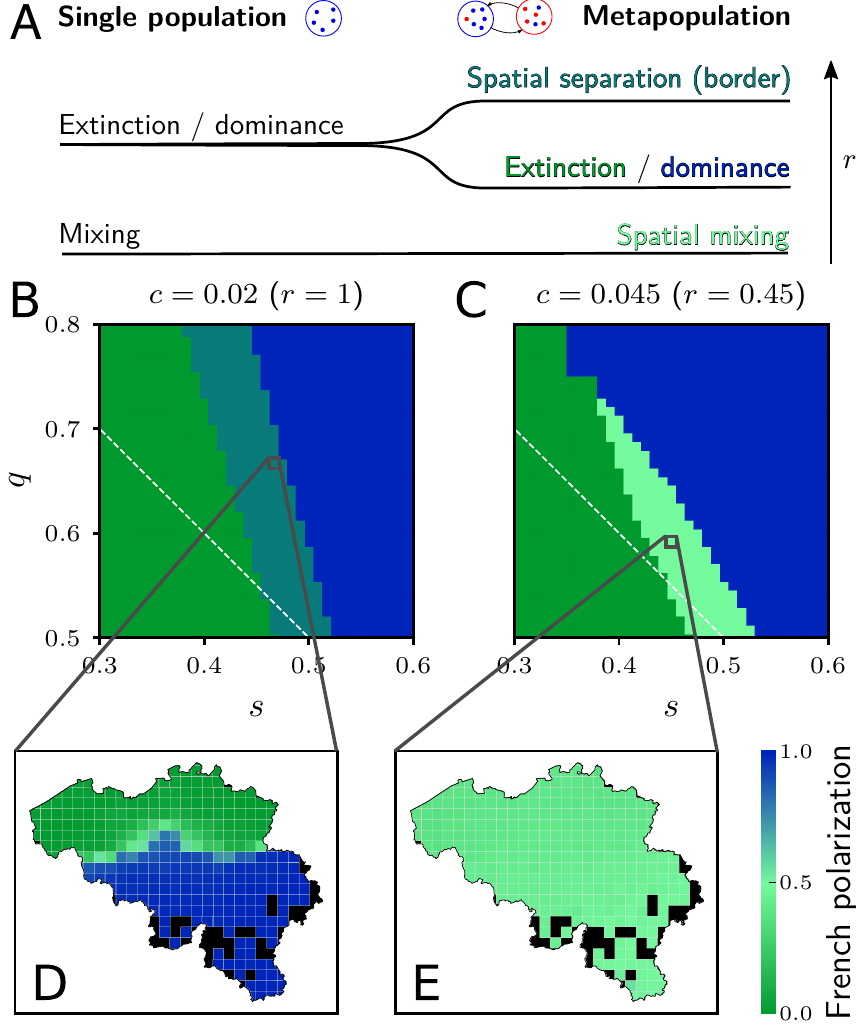}
    \caption{Types of stable states of convergence of our model in a metapopulation set for Belgium. (\textit{A}) Diagram illustrating the effect of adding metapopulations in the stable states of a single population: the former extinction state bifurcates in full extinction and in a boundary-like state with monolinguals separated in space. Larger values of $r$ favor homogeneity, either by full extinction or by separation states (see the SI Fig. S16 for more $r$ values \cite{supp}). Below are the regions of the parameter space $(s,q)$ where these stable solutions emerge, (\textit{B}) with $r=1$ and (\textit{C}) with $r=0.45$. Finally, two polarization maps show examples of states the model converges to, (\textit{D}) a boundary-like state for $r=1$, $s=0.467$, $q=0.667$, and (\textit{E}) complete mixing for $r=0.45$, $s=0.45$, $q=0.592$.}
    \label{fig:single_to_metapop_BE}
\end{figure}

The populations and commuting are thus obtained from the national census at municipality scale (see Methods for how to access them). We implement a mapping process from municipalities to our cells based on area overlap (details in the SI Sec. IV A \cite{supp}). Regarding the language groups, Twitter data may suffer from different socio-demographic biases \cite{Mislove2011,Nguyen2015}, and besides tweets reflect language use online, not necessarily the offline practices in the full population. Since in the census we found information on the total number of persons per language group and of residents per municipality, we have scaled the $L$-speakers that we find on Twitter to match these two sets of marginal sums via Iterative Proportional Fitting (IPF) \cite{Deming1940,Fienberg1970}.

Once the metapopulation has been initialized, the model can be simulated. As in Ref. \cite{Fernandez-Gracia2014}, the day is divided in two parts: the individuals first start in their residence cells and interact with the local agents following the rates of equation \eqref{eq:bipref_model}, and then move to their work cells where again they interact with the local population. The agents encounter thus different environments characterized by diverse proportions $p_{L,i}$ in the two parts of the day. Even if they live and work in the same cell, the local population changes from one part of the day to the next.

In order to analyze the stability of the steady states reached by the extended model, we derive an approximate master equation for the full metapopulation setting. To this end, we adapt the methodology described in Refs. \cite{Sattenspiel1995,Balcan2010} for epidemiological models (see the SI Sec. IV B for details \cite{supp}). The equations obtained are only approximated but since they are analytic we can integrate them and calculate the Jacobian at their fixed points. To check the consistency of both approaches and that the fixed points of the dynamics are the same, we also introduce the initial conditions in the master equation, to then integrate it numerically using a standard Runge-Kutta algorithm. The fixed points reached by the simulations turn out to be fixed points as well for the equations. Not only that, all the eigenvalues of the Jacobian at these states have negative real parts and they are thus stable fixed points.

To explore the parameter space systematically, we perform a number of simulations until convergence to a stable state. We show the results for the metapopulation setting of Belgium in Fig. \ref{fig:single_to_metapop_BE}. 
Remarkably, a new kind of stable state emerges. While in a single population we had only two stable configurations: extinction or mixing, here we can find full mixing (Fig. \ref{fig:single_to_metapop_BE}\textit{E}), global extinction and local extinction of a language in part of the territory leading to a boundary-like state (Fig. \ref{fig:single_to_metapop_BE}\textit{D}). This state of convergence is similar to the initial conditions, corresponding to the language border we observe today. We have thus checked that our model, in these conditions, is able to obtain the present state as a stable solution. A surprising aspect of the results is that decreasing $r$, or in other words making it easier or more common to learn the other language, does not necessarily favor coexistence. Indeed, as $r$ decreases, at one point boundary states become unstable and this may not necessarily lead to fully mixed states. When $r$ shrinks bilinguals become more numerous on the boundary, until they expand beyond the boundary and spread bilingualism across the region. Still, if this happens when $r$ is not low enough, the two languages cannot coexist and one ends up extinct, as the coexistence region of the parameter space in a single population shown in Fig. \ref{fig:coex_region} may not have been reached. 

We also wished to explore the possibility of having a hybrid state, consisting in an area where a minority language survives through bilinguals within an otherwise monolingual region. This is the case of Sundanese and Javanese in Java for instance (see SI Fig. S7 \cite{supp}). We initialized a hypothetical population in Belgium, with only monolinguals in Dutch, except in a pocket of cells in the South of the country, where there are only bilinguals. The latter were attached a $q = 0.62$, while $q = 0.5$ for the rest. Iterating the model yields a stable solution similar to this initial state, with a mix of bilinguals and Dutch monolinguals in the pocket, and only Dutch monolinguals elsewhere (see SI Fig. S15 \cite{supp}).

\subsection{Dynamics in the parameters}
\begin{figure}
\centering
    \includegraphics[width=\linewidth]{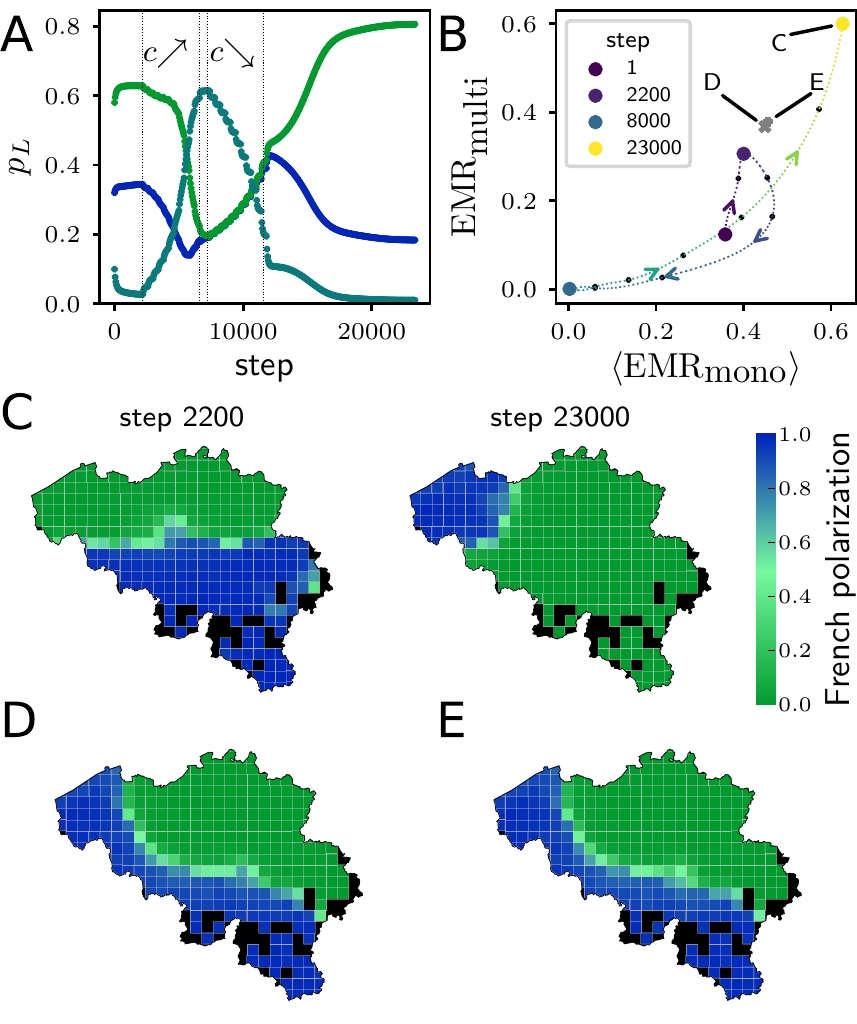}
    \caption{Evolution of the state of the metapopulation model in Belgium when $c$ varies, first slowly increased and then decreased to recover the original value. We fixed $s=q=1/2$ and $\mu=0.02$. (\textit{A}) Evolution of the global proportions $p_L$ of individuals belonging to each $L$-group. The blue curve corresponds to French monolinguals, light green to Dutch monolinguals and dark green to bilinguals. (\textit{B}) Trajectory of the system in the EMR space: on the $x$-axis the average of the EMR between each monolingual community and the whole population, and on the $y$-axis the one between bilinguals an the whole population. The initial state and the stable states the system went through are marked by colored circles, while black ones mark additional points where the EMR was calculated, and the dashed line the interpolation between them. (\textit{C}) Polarization maps of French in the initial and final states, both featuring a boundary but located in different areas, thus showing the irreversibility of the dynamics. (\textit{D})-(\textit{E}) Polarization maps of French in the final states of simulations including trans-border commuters from France and the Netherlands, respectively with proportions $p_{\text{TB}}$ equal to $0.5\%$ and $0.2\%$ of the population of the border municipalities of these two countries. The points in the EMR space corresponding to these final states are also represented in panel (\textit{B}).}
    \label{fig:varying_c_sep_to_mix_to_sep}
\end{figure}
    
The effect of multilingual education or, in general, policies favoring the use of one or several languages can alter the values of our model parameters. For example, $c$ represents how monolinguals learn the other language. This process can be facilitated by the similarity between the languages or by teaching in both languages at school, for instance. Next, we investigate whether a parameter changing in time can perturb the system out of a stable state, and how the transition to a completely different configuration occurs. To this end, we run a simulation for 23000 steps and present the results in Fig. \ref{fig:varying_c_sep_to_mix_to_sep}. To explore the effects of the $c$ parameter evolution alone, we fix the other parameters $s = q = 1/2$ and $\mu = 0.02$. We start from our initial conditions with $c = 0.005$, which converges to a stable state with a boundary (see the first map of Fig. \ref{fig:varying_c_sep_to_mix_to_sep}\textit{C}). After 2200 steps, we then increase $c$ by $0.005$ every $400$ steps until we reach $c = 0.055$. The system converges quickly to a state of mixed coexistence, with a majority of bilinguals and equal proportions of monolinguals, like in Fig. \ref{fig:single_to_metapop_BE}\textit{E}. $c$ is then decreased at the same rate as before to reach its initial value of $0.005$. The system eventually converges to a state displaying a boundary, but displaced compared to its initial position. A visualization of this evolution is proposed in movie S1. Also, the resulting trajectory in the EMR space in Fig. \ref{fig:varying_c_sep_to_mix_to_sep}\textit{B} shows that the final stable state exhibits more segregation for both monolinguals and bilinguals, since the boundary between communities lies in the countryside, and not around Brussels as in the original scenario. The importance of the history of languages is hence clearly shown by this experiment.

The seemingly random placement of the boundary may be owed to the absence of constraints on the system, which is completely closed. In reality a country is an open system with exterior influences, notably from its direct neighbors.
Thus, we ran the same simulation with trans-border proportions $p_{\text{TB}}$ equal to $0.5\%$ and $0.2\%$ of the population of the border municipalities of France and the Netherlands commuting to Belgium. These commuters act as a fixed population of monolinguals interacting only during the workday with the local population (for more details, see SI Sec. IV D \cite{supp}). These boundary conditions stabilize the final state of convergence, as the linguistic boundary resulting from the process of varying $c$ is similar for the two values of $p_{\text{TB}}$, following the orientation of the two opposite borders (see Fig. \ref{fig:varying_c_sep_to_mix_to_sep}\textit{D}-\textit{E}). This positioning is a clear improvement over the closed-system simulation, albeit still not quite the one we observed in Fig. \ref{fig:BE_CAT_maps_EMD_scatter}. In Fig. \ref{fig:varying_c_sep_to_mix_to_sep}\textit{B}, the positions of these two states in the EMR space are also shown to be much closer to the original state than the final state of the first trajectory.

More complex settings could be envisaged to get closer to a realistic solution. A space-dependent prestige could be introduced, taking different values in Flanders, Wallonia and Brussels for instance. Also, we here considered only the commuting part of human mobility, but other kinds of mobility like migrations may have their importance. This is especially true for attractive metropolises like Brussels, which are typically places of intense language contact \cite{Simon2011}.
However, in this simulation the aim was to check the irreversibility of a change when increasing the ease to learn the other language and subsequently decreasing it to its original value, which was indeed confirmed.

\section{Discussion}
In summary, we have explored the spatial distribution patterns of language competition and coexistence in multilingual societies. We first did so by introducing the Earth Mover's Ratio, a metric capable of measuring the spatial segregation of a group in a given society, starting from a distance between its distribution and that of the whole population. Two main configurations have thus been observed: either spatial mixing with multilinguals widespread, or separate linguistic groups with a clear boundary between them and multilinguals concentrating around it. 

Despite the ubiquity of these two configurations and their apparent temporal stability, the models introduced in the literature were not able to offer clear solutions capturing them. As we show, the main difficulty comes from the role of bilinguals in keeping languages alive. In many occasions, the monolingual community of one of the languages may become virtually extinct and its use relies only on the bilingual group. We have introduced a model taking this into account and have shown that it is able to produce naturally both configurations as stable solutions without the need for artificial nonlinearities. The model features a parameter considering the preference of bilinguals for one of the two languages. This preference actually acts as a kind of defense mechanism since the use by bilinguals of the endangered language may be enough to save it, countering a possibly lower prestige of the language within society as a whole. The ease to learn the other language also has a role in the model. It may be influenced by both the similarity between languages, which can hardly be controlled, but also by the policies put into place to facilitate its learning. We have shown that this parameter is critical to determine whether languages can coexist. The parameters of the model could be estimated using longitudinal data. The scope of this work was not predictive, but rather to study stable solutions of the model, so we leave it here for future work.

When spatial interactions are taken into account via the commuting patterns of individuals, the model is able to reach a stable state where two language communities are separated by a boundary around which they coexist. In this case, however, we have shown that, quite counter-intuitively, increasing this ease to learn the other language may break the existing boundary and lead to extinction, and not to the desired coexistence with mixing of the languages. This calls for caution when designing policies since the final state is strongly history-dependent. %Encouraging bilinguals to prefer their minority language  

Overall, our findings shed light on the role of heterogeneous speech communities in multilingual societies, and they may help shape the objectives and nature of language planning \cite{Kaplan1997} in many countries where accelerated changes are threatening cultural diversity.

\section{Materials and Methods}
To facilitate the reproducibility of our research, we provide here the direct sources of the data we used, when possible, and the method to acquire them when they are subject to privacy concerns. The code developed for this work is also made available in a public repository.

\subsection{Data access}
The geo-located tweets used to map language use were collected through the streaming API of Twitter, and more specifically using the ``statuses/filter'' endpoint: \href{https://developer.twitter.com/en/docs/twitter-api/v1/tweets/filter-realtime/overview}{https://developer.twitter.com/en/docs/twitter-api/v1/tweets/filter-realtime/overview}. This endpoint provides a sample of tweets in real time matching some provided filters. For the purpose of this work, bounding box filters were set to collect tweets from a set of countries of interest. Before reproducing this method of data collection, one should bear in mind that the current form and even the availability of this endpoint is subject to future changes introduced by the Twitter Developer's team. The aggregated data giving the counts of local users by language group by cell have been deposited on figshare (\href{https://figshare.com/articles/dataset/Spatial_distributions_of_languages_on_Twitter/14339321}{https://figshare.com/articles/dataset/Spatial\_\\distributions\_of\_languages\_on\_Twitter/14339321}).

The data on commuting patterns at the municipality level in Belgium were obtained from the 2011 census (available for download at \href{https://statbel.fgov.be/en/open-data/census-2011-matrix-commutes-sex}{https://statbel.fgov.be/en/open-data/census-2011-matrix-commutes-sex}). The data about the knowledge of official languages (English or French or both) by census subdivisions in Quebec were obtained from the 2016 Canadian census, and can be downloaded directly from \href{https://www12.statcan.gc.ca/census-recensement/2016/dp-pd/dt-td/Rp-eng.cfm?TABID=2&LANG=E&A=R&APATH=3&DETAIL=0&DIM=0&FL=A&FREE=0&GC=01&GL=-1&GID=1159582&GK=1&GRP=1&O=D&PID=110461&PRID=10&PTYPE=109445&S=0&SHOWALL=0&SUB=0&Temporal=2016&THEME=118&VID=0&VNAMEE=&VNAMEF=&D1=0&D2=0&D3=0&D4=0&D5=0&D6=0}{https://www12.statcan.gc.ca/census-recensement/2016/dp-pd/dt-td/Index-eng.cfm}.

\subsection{Code}
The data processing, the plotting of results and the simulations were carried out in Python with the help of open-source libraries. All of the Python code used for this work is hosted on GitHub: \href{https://github.com/TLouf/multiling-twitter}{https://github.com/TLouf/multiling-twitter}. Mathematica was used to carry out part of the analytic work on the models and to generate the associated Figs. \ref{fig:stream_plots} and \ref{fig:coex_region}. The corresponding code is also hosted on GitHub, available at \href{https://github.com/TLouf/multiling-analytical}{https://github.com/TLouf/multiling-analytical}.

\begin{acknowledgments}
The authors acknowledge funding from the project PACSS (RTI2018-093732-B-C22) of the MCIN/AEI/10.13039/501100011033/ and of the EU through FEDER funds (A way to make Europe) and also from MCIN/AEI/10.13039/501100011033/ under the Maria de Maeztu program for Units of Excellence in R\&D (MDM-2017-0711). This work has been carried out within the COST Action Nexus Linguarum CA 18209. 
\end{acknowledgments}

% Bibliography
\bibliography{biblio}

\end{document}

\title{Supplementary Information \\Capturing the diversity of multilingual societies}
\author{Thomas Louf}
\email{thomaslouf@ifisc.uib-csic.es}
\author{David S\'{a}nchez}
\author{José J. Ramasco}
\affiliation{Institute for Cross-Disciplinary Physics and Complex Systems IFISC (UIB-CSIC), 07122 Palma de Mallorca, Spain}
\maketitle

\section{Twitter data analysis}
\label{sec:data_analysis}
\subsection{Data cleaning}
As this work is solely concerned with the language use of the locals of the regions considered, we started the analysis by filtering out users whose behavior did not fit this profile. We first eliminate bots tweeting at an inhuman rate, set at an average of three tweets per minute between the first and last tweet of a user. Consecutive geo-locations implying speeds higher than a plane's (\SI{1000}{\kilo \meter \per \hour}) are also detected to discard users. The final filter aims at keeping only residents of the region considered, as it imposes for a user to tweet from there in at least three consecutive months.

\subsection{Residence attribution}
\label{sec:res_attrib}
The next step is to attribute the remaining users to a cell of residence in a regular grid drawn over the region of interest, which defines the bins of the spatial distributions of languages. 
Before 2015, Twitter made available the exact coordinates of where a tweet was sent from when the user activated geo-location services. However, that year they changed their privacy policy and since then, by default, a user who enabled geo-location will not get their coordinates attached to their tweets, only a ``place'' is assigned to each tweet. A place is an area with a bounding box, which can have different scales:
\begin{itemize}
    \item country,
    \item administrative unit: province, region or department for instance,
    \item city,
    \item place of interest (POI): any kind of public place: restaurant, school, event venue, etc. These are represented by a point, so we considered tweets attached to a POI similarly to the ones having coordinates attached.
\end{itemize}
When a user tweets with their device's GPS activated, a place (usually the city they tweet from) is selected by default, and they can switch to another one from a list of close-by places. These places are fed to Twitter by Foursquare \cite{Foursquare}, which provides data down to a POI level for more than 190 countries. Users can still activate exact GPS coordinates for their tweets on their profile, however as it is not the default setting, only a small proportion of them make the change (about 10 to 20\%, depending on the country). Thus, when the coverage of Foursquare is lacking in a country, the amount of location data available to find the residence of users is greatly reduced, and sometimes not enough to gather sufficient statistics. Bolivia for instance was one such case.

For every user, from a collection of tweets with either GPS coordinates or a bounding box as geo-location, we need to assign a cell of residence. First, for every tweet with exact GPS coordinates, we determine the cell these belong to. We only keep the cells where at least 10\% of their tweets originated from, if any, and take the one from which they tweeted the most outside work hours. This way, we know that this cell is relevant and not just somewhere the user occasionally travels to, and also that when they are most likely at home, it is where they tweet the most often from. For the users who did not have any cell above the relevance threshold, we then proceed to find their place of residence. This happens for users who did not activate exact location tracking and who do not tweet from many POIs, either because they do not bother to select one when tweeting, or because they live in an area where there are few or none of them. The latter happens quite often in small towns. We only consider places which are smaller than a certain area, which depends on each country's size and places' data. A typical value of this threshold is \SI{1000}{\kilo \meter \squared}. This is to discard places which would not actually give much location information, since they cover a significant part of the region of interest. If a user only tweets from places which are too large, we discard them completely, as in this case we do not have any accurate enough data on their residence. Then, like we did for cells, we pick the relevant place with the most tweets outside work hours. There remains to translate this place of residence into the previously-defined grid of cells. Since a place can intersect multiple cells, we attribute users to all intersected cells, proportionally to the portion of the area of the place contained within each cell.

\subsection{Language attribution}
\label{sec:lang_attrib}
The final step required to obtain the desired spatial distributions consists in attributing language(s) to each user. Our starting point is the text of all their tweets. To detect the language of each of these, we used Chromium's Compact Language Detector (CLD) \cite{Solomon2014}, which from an input text returns the most probable language it was written in along with a confidence. The tweets' text needed to be cleaned beforehand though, as it may contain URLs, hashtags, mentions and even text generated by third party applications, all of which are not relevant to detect the language spoken by the user. Once a tweet has been stripped of all such elements, we ensure that it remains long enough (4 words) to retain a reliable detection. All tweets exceeding this threshold are then passed through the CLD, and the result is considered if it returned a confidence above $90\%$. We then obtain the counts of tweets each user sent by language. As a user may occasionally tweet in a language they do not speak by way of quoting someone else or using a translator, we do not keep all of them. We set that at least $10\%$ or 5 of the tweets of a user must be in a certain language to consider them a speaker in this language. At the end of this data processing, every user deemed resident of the region of interest has been attributed a cell of residence $i$ and a set of languages.

\subsection{Measuring spatial segregation}
Building upon proportion or concentration of speakers, we want to be able to measure the spatial mixing of language groups, or inversely, their spatial segregation. We define segregation as the difference in how individuals of a given group are spatially distributed compared to the whole population of the area. Segregation is thus conceptualized as the departure from a null model, the unsegregated scenario, in which regardless of the group an individual belongs to, they would be distributed randomly according to the whole population's distribution. Explicitly, the concentrations corresponding to the null models are $c_i = N_i / N$. To quantify language mixing, we measure a distance between the spatial distribution of a given language group and that of the whole population.

\begin{figure}[p]
  \centering
  \includegraphics{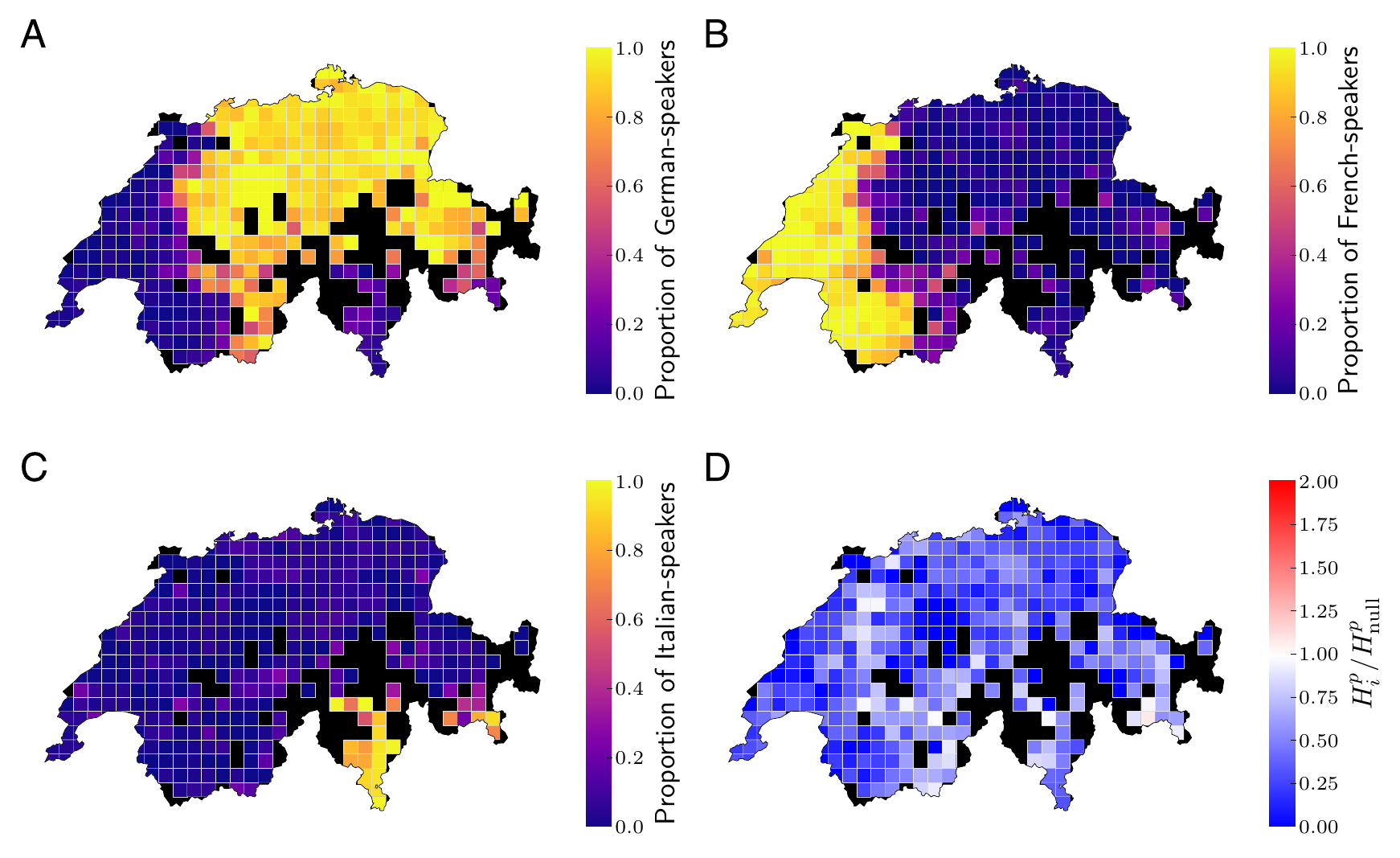}
  \caption{Mapping of the local languages in Switzerland from Twitter data. For each cell of $10 \times 10 \, \si{\kilo \meter \squared}$, the proportions $p_{l,i}$ of (\textit{A}) German, (\textit{B}) French and (\textit{C}) Italian speakers are shown, as well as (\textit{D}) the relative proportion entropy of the $L$ groups. In black are cells in which fewer than 5 Twitter users speaking a local language were found to reside, consequently discarded for the insufficient statistics. A clear separation of language groups is visible in Switzerland following the linguistic regions, displaying mixing mainly around the border between the German and French speaking regions.}
  \label{fig:twitter_data_CH}
\end{figure}

\begin{figure}[p]
  \centering
  \includegraphics{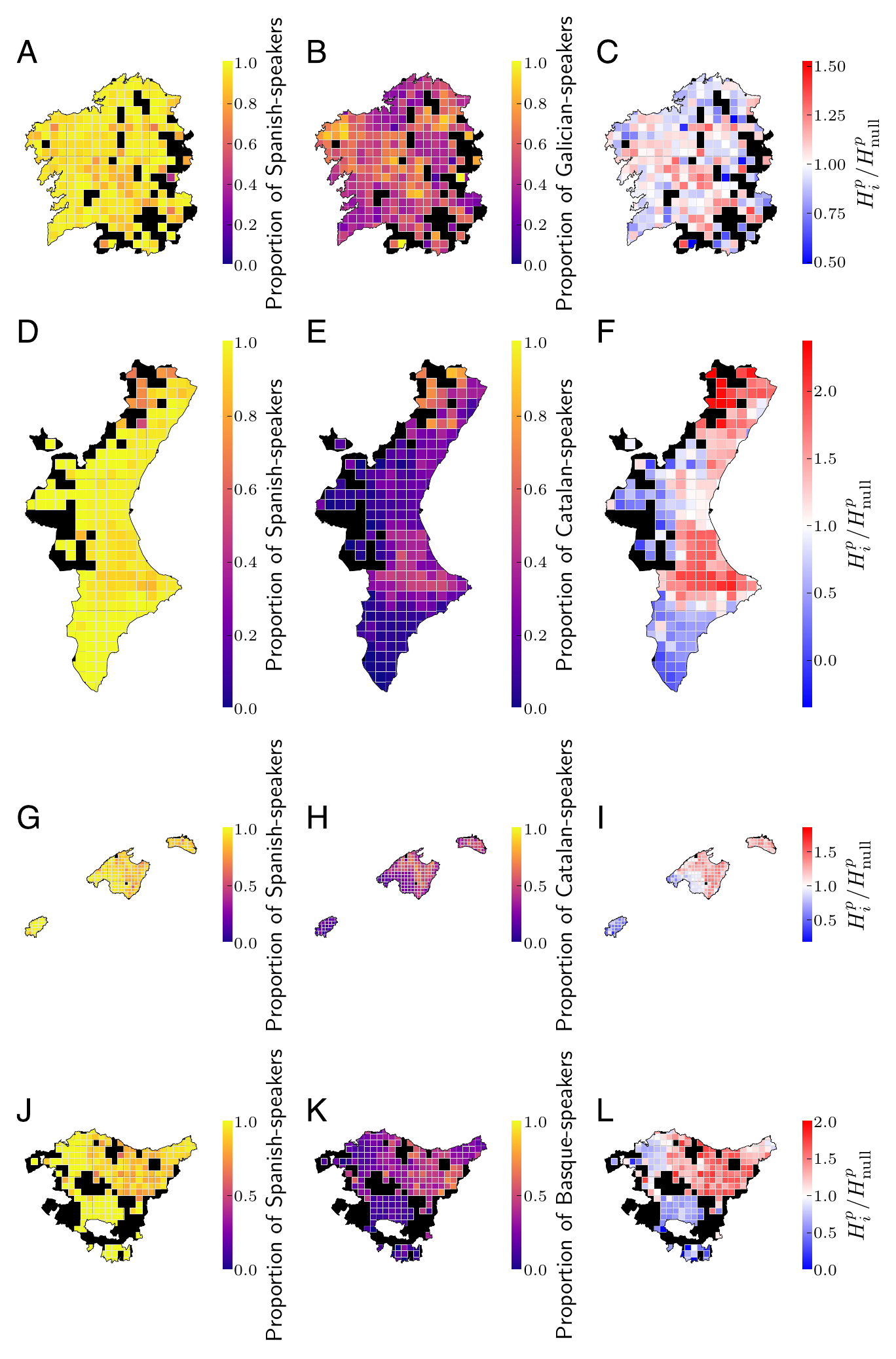}
  \caption{Mapping of the local languages in several autonomous communities of Spain from Twitter data. For each cell of $10 \times 10 \, \si{\kilo \meter \squared}$, the proportions $p_{l,i}$ of (\textit{A}) Spanish and (\textit{B}) Galician speakers are shown in Galicia, as well as (\textit{C}) the relative proportion entropy of the $L$ groups. In the Valencian Community are mapped the proportions of (\textit{D}) Spanish and (\textit{E}) Catalan speakers, as well as (\textit{F}) the relative proportion entropy of the $L$ groups. For cells of $5 \times 5 \, \si{\kilo \meter \squared}$, in the Balearic Islands are mapped the proportions of (\textit{G}) Spanish and (\textit{H}) Catalan speakers, as well as (\textit{I}) the relative proportion entropy of the $L$ groups. In the Basque Country are mapped the proportions of (\textit{J}) Spanish and (\textit{K}) Basque speakers, as well as (\textit{L}) the relative proportion entropy of the $L$ groups. In black are cells in which fewer than 5 Twitter users speaking a local language were found to reside, consequently discarded for the insufficient statistics. The use of Spanish is widespread, while the use of the language specific to each region is more present in the countryside.}
  \label{fig:twitter_data_add_ES}
\end{figure}

\begin{figure}[p]
  \centering
  \includegraphics{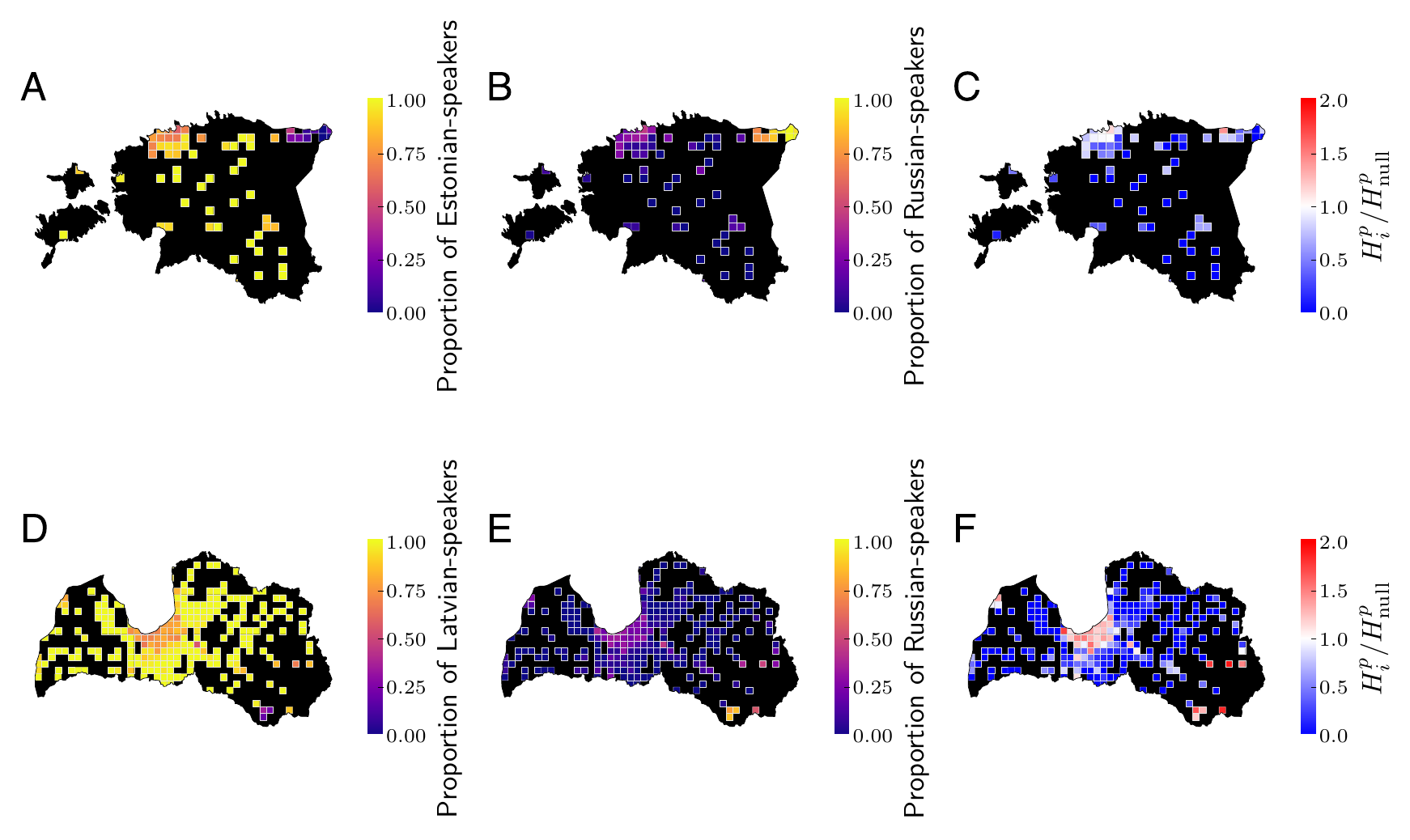}
  \caption{Mapping of the local languages in Estonia and Latvia from Twitter data. For each cell of $10 \times 10 \, \si{\kilo \meter \squared}$, the proportions $p_{l,i}$ of (\textit{A}) Estonian and (\textit{B}) Russian speakers are shown in Estonia, as well as (\textit{C}) the relative proportion entropy of the $L$ groups. In Latvia are mapped the proportions $p_{l,i}$ of (\textit{D}) Latvian and (\textit{E}) Russian speakers, as well as (\textit{F}) the relative proportion entropy of the $L$ groups. In black are cells in which fewer than 5 Twitter users speaking a local language were found to reside, consequently discarded for the insufficient statistics. In the two countries, the mixing the the two languages mainly occur in the capital city, while Russian is barely present in the countryside. The main difference between the two is that in Estonia, a rather large city on the border with Russia is dominated by Russian speakers.}
  \label{fig:twitter_data_EE_LV}
\end{figure}

\begin{figure}[p]
  \centering
  \includegraphics{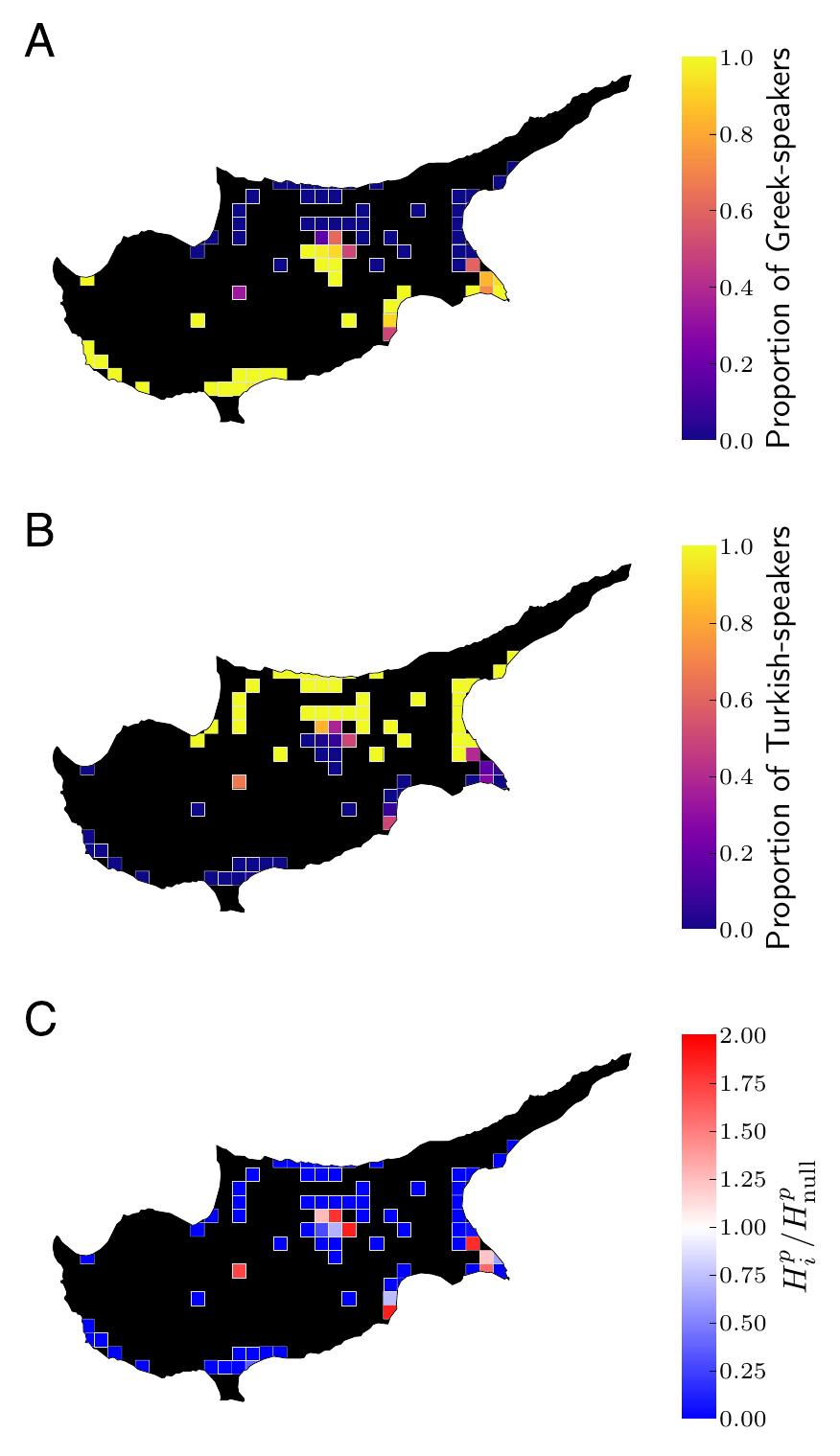}
  \caption{Mapping of the local languages in Cyprus from Twitter data. For each cell of $5 \times 5 \, \si{\kilo \meter \squared}$, the proportions $p_{l,i}$ of (\textit{A}) Greek and (\textit{B}) Turkish speakers are shown, as well as (\textit{C}) the relative proportion entropy of the $L$ groups. In black are cells in which fewer than 5 Twitter users speaking a local language were found to reside, consequently discarded for the insufficient statistics. A clear North-South separation is visible.}
  \label{fig:twitter_data_CY}
\end{figure}

\begin{figure}[p]
  \centering
  \includegraphics{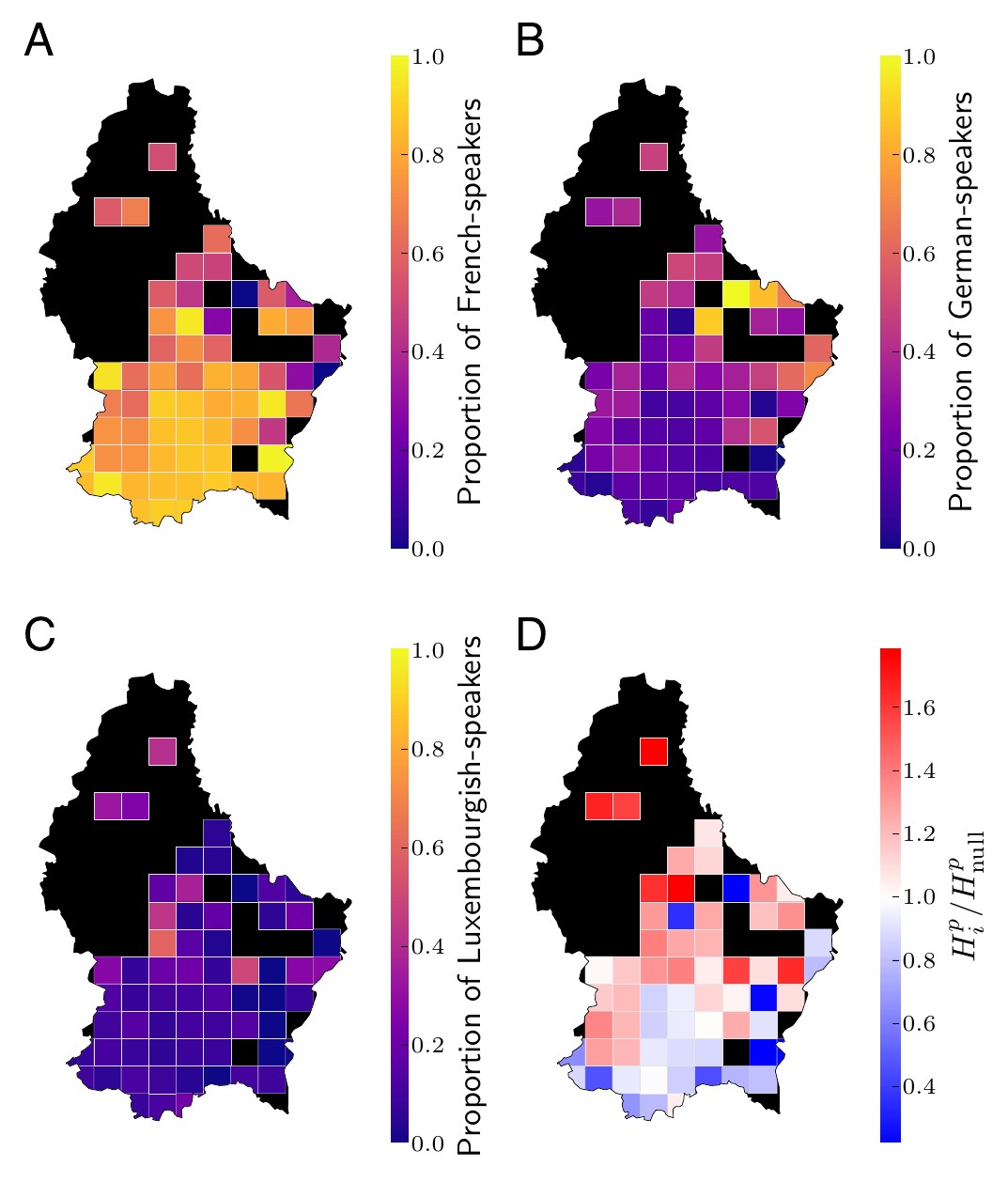}
  \caption{Mapping of the local languages in Luxembourg from Twitter data. For each cell of $5 \times 5 \, \si{\kilo \meter \squared}$, the proportions $p_{l,i}$ of (\textit{A}) French, (\textit{B}) Dutch and (\textit{C}) Luxembourguish speakers are shown, as well as (\textit{D}) the relative proportion entropy of the $L$ groups. In black are cells in which fewer than 5 Twitter users speaking a local language were found to reside, consequently discarded for the insufficient statistics. While French-speakers are dominant around the capital city, German-speakers are more present in the North-East, close to the border with Germany. Luxembourgish is hardly visible on Twitter.}
  \label{fig:twitter_data_LU}
\end{figure}

\begin{figure}[p]
  \centering
  \includegraphics{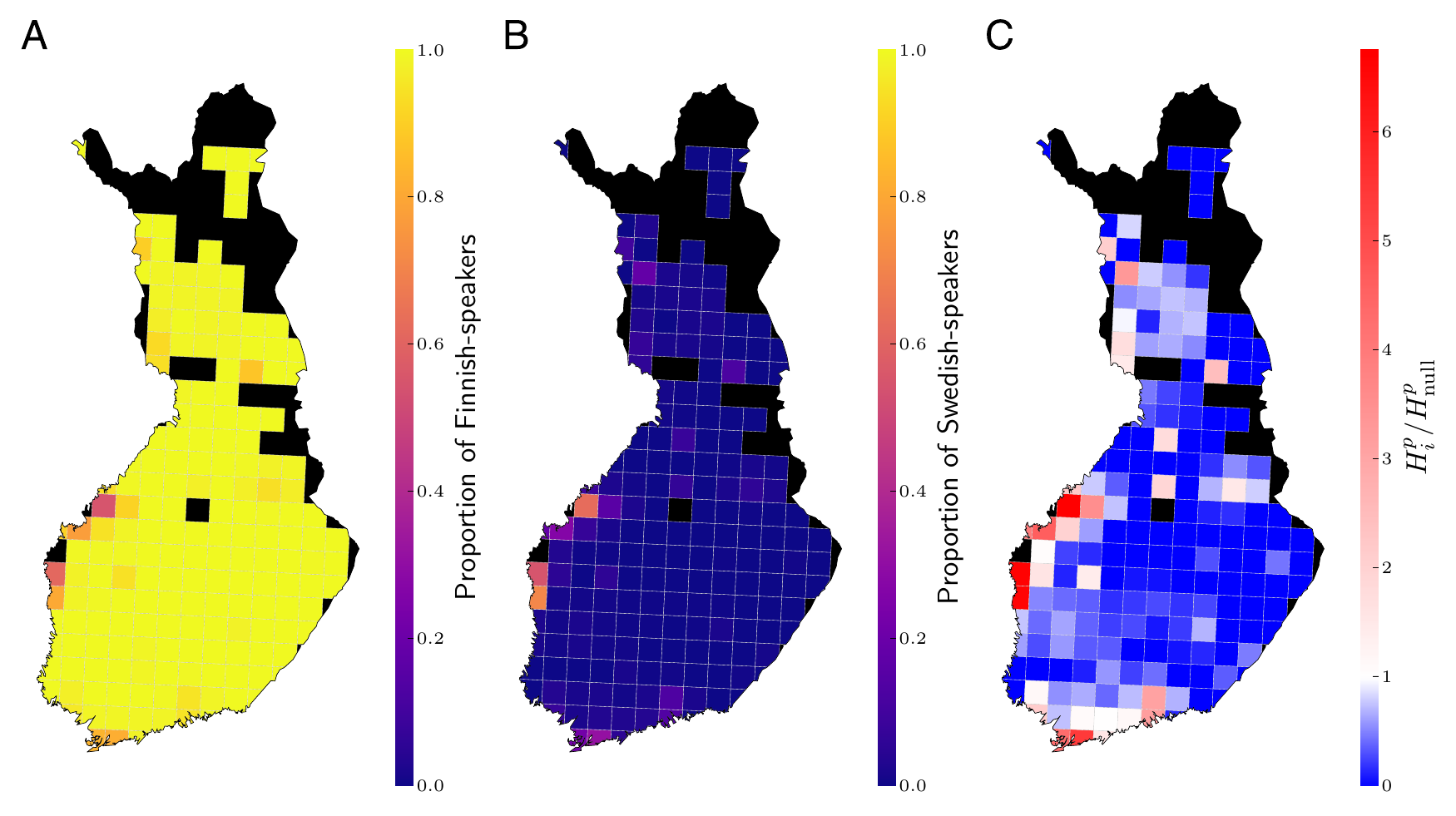}
  \caption{Mapping of the local languages in Finland from Twitter data. For each cell of $40 \times 40 \, \si{\kilo \meter \squared}$, the proportions $p_{l,i}$ of (\textit{A}) Finnish and (\textit{B}) Swedish speakers are shown, as well as (\textit{C}) the relative proportion entropy of the $L$ groups. In black are cells in which fewer than 5 Twitter users speaking a local language were found to reside, consequently discarded for the insufficient statistics. A clear separation is visible. Finnish speakers are in great majority, but some strong isolated communities of Swedish speakers are visible, mostly in the West.}
  \label{fig:twitter_data_FI}
\end{figure}

\begin{figure}[p]
  \centering
  \includegraphics{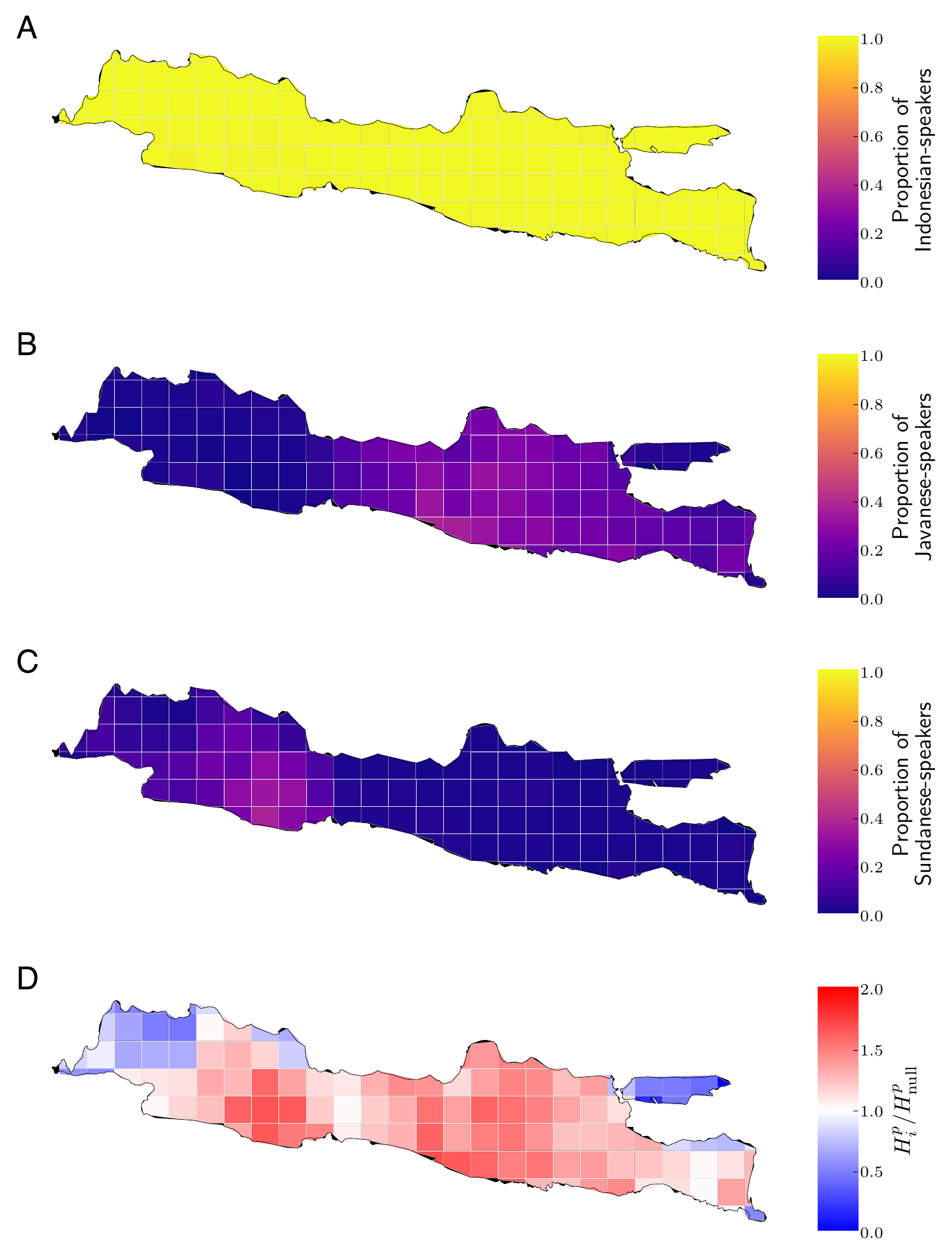}
  \caption{Mapping of the local languages on the Indonesian island of Java from Twitter data. For each cell of $40 \times 40 \, \si{\kilo \meter \squared}$, the proportions $p_{l,i}$ of (\textit{A}) Indonesian, (\textit{B}) Javanese and (\textit{C}) Sundanese speakers are shown, as well as (\textit{D}) the relative proportion entropy of the $L$ groups. In black are cells in which fewer than 5 Twitter users speaking a local language were found to reside, consequently discarded for the insufficient statistics. Interestingly, the Javanese and Sundanese speaking communities are very localized  but mix with the majority of Indonesian speakers in their regions, as most of the speakers of the two former languages are bilingual.}
  \label{fig:twitter_data_Java}
\end{figure}

\begin{figure}[p]
  \centering
  \includegraphics{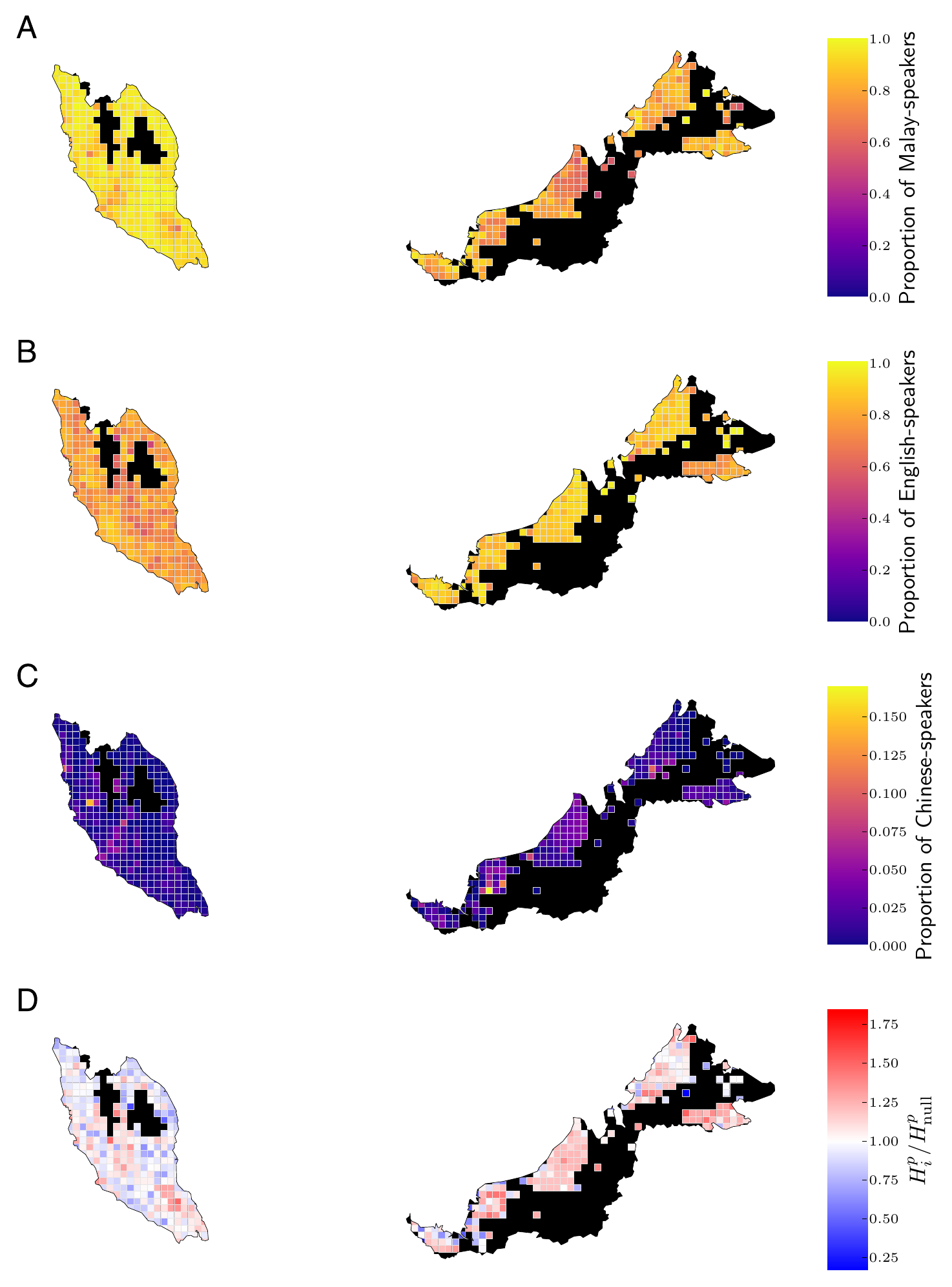}
  \caption{Mapping of the local languages on Malaysia from Twitter data. For each cell of $20 \times 20 \, \si{\kilo \meter \squared}$, the proportions $p_{l,i}$ of (\textit{A}) Malay, (\textit{B}) English and (\textit{C}) Chinese speakers are shown, as well as (\textit{D}) the relative proportion entropy of the $L$ groups. In black are cells in which fewer than 5 Twitter users speaking a local language were found to reside, consequently discarded for the insufficient statistics. Malay and English are both widespread while Chinese is mainly spoken around the largest urban areas.}
  \label{fig:twitter_data_MY}
\end{figure}

\begin{figure}[p]
  \centering
  \includegraphics{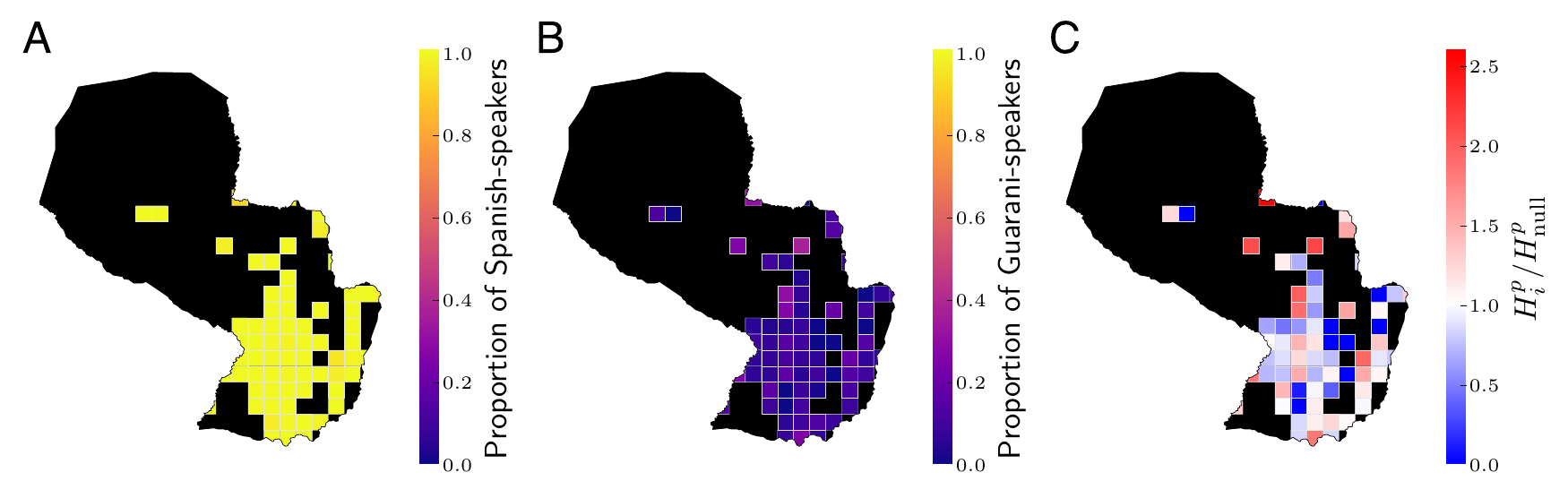}
  \caption{Mapping of the local languages in Paraguay from Twitter data. For each cell of $40 \times 40 \, \si{\kilo \meter \squared}$, the proportions $p_{l,i}$ of (\textit{A}) Spanish and (\textit{B}) Guarani speakers are shown, as well as (\textit{C}) the relative proportion entropy of the $L$ groups. In black are cells in which fewer than 5 Twitter users speaking a local language were found to reside, consequently discarded for the insufficient statistics. Spanish is spoken everywhere while Guarani's usage is more sparse.}
  \label{fig:twitter_data_PY}
\end{figure}

Metrics based on the principle of entropy are among the most popular options to measure segregation~\cite{Reardon2002a}. An entropy can thus be defined over the proportion distributions over language groups $L$ in every cell $i$ as
\begin{equation}
    \label{eq:def_Hp}
    H_i^p = - \sum_{L} p_{L, i} \log (p_{L, i}).
\end{equation}
To actually measure mixing in cell $i$, this needs to be compared to its value in the null model, the unsegregated scenario. The former gives the following proportion entropy:
\begin{equation}
    H_{\text{null}}^p = - \sum_{L} \frac{N_{L}}{N} \log \left( \frac{N_{L}}{N} \right).
\end{equation}
If the ratio $H_i^p / H_{\text{null}}^p$ is below $1$, cell $i$ is less mixed than the region, which means that the majority groups are even more in majority and the minorities even more in minority than they should. If it is above $1$, the groups are more evenly distributed, but still segregated, as this is brought by a minority group being over-represented, which is the case when a minority is concentrated in ghettos. What is interesting to us then is simply how much we deviate from $1$, a value reached when cell $i$'s mixing is identical to the region's. A visualization of this relative entropy in the regions of interest listed in Table \ref{tab:region_counts} is proposed in Figs. \ref{fig:twitter_data_CH}-\ref{fig:twitter_data_PY}. These figures also show the proportion of individuals speaking each language of the region.

This ratio of proportion entropy enables the visualization of the mixing of language groups in space, but we would also like to define a global index quantifying how segregated each language group is. An entropy over the distribution of concentrations over all cells $i$ could be used to that end, as was done in \cite{Lamanna2018}. To do so, as this distribution is spatial, a tweak would be required to properly take into account the size of bins, as shown in \cite{Batty1974}.
However, even with this tweak this kind of entropy still does not properly take into account the spatial embedding of the distributions. This was thoroughly discussed in the segregation literature \cite{White1986,Reardon2002,Dawkins2007} and illustrated by the checkerboard problem. This problem arises from the simple fact that if one was to move around the cells of the system, thus potentially forming (breaking) clusters, the entropies would remain completely unchanged, while the value of the segregation metric should increase (decrease). Hence the need for another metric introduced in the main text: the Earth Mover's Ratio (EMR). For every $L$-group corresponding to languages which are local of our regions of interest and for which sufficient statistics were gathered from Twitter, we give the corresponding EMR in Table \ref{tab:region_summary}.

\subsection{Cell sizes}
\label{sec:cell_sizes}
The choice of the size of the cells is critical, as it defines the bins of the spatial distributions that we are studying. There are two limits to the cell size. First is an upper one, because if too many users are aggregated, we simply cannot see segregation anymore. The second is a lower one due to the nature of the geo-location data at hand: the Twitter places data described above in Sec. \ref{sec:res_attrib}. The scale of the places data varies between countries, as for instance cities in America are typically more extended than in Europe. Considering these two constraints, we thus chose cell sizes carefully for each region considered.

\begin{figure}[p]
  \centering
  \includegraphics{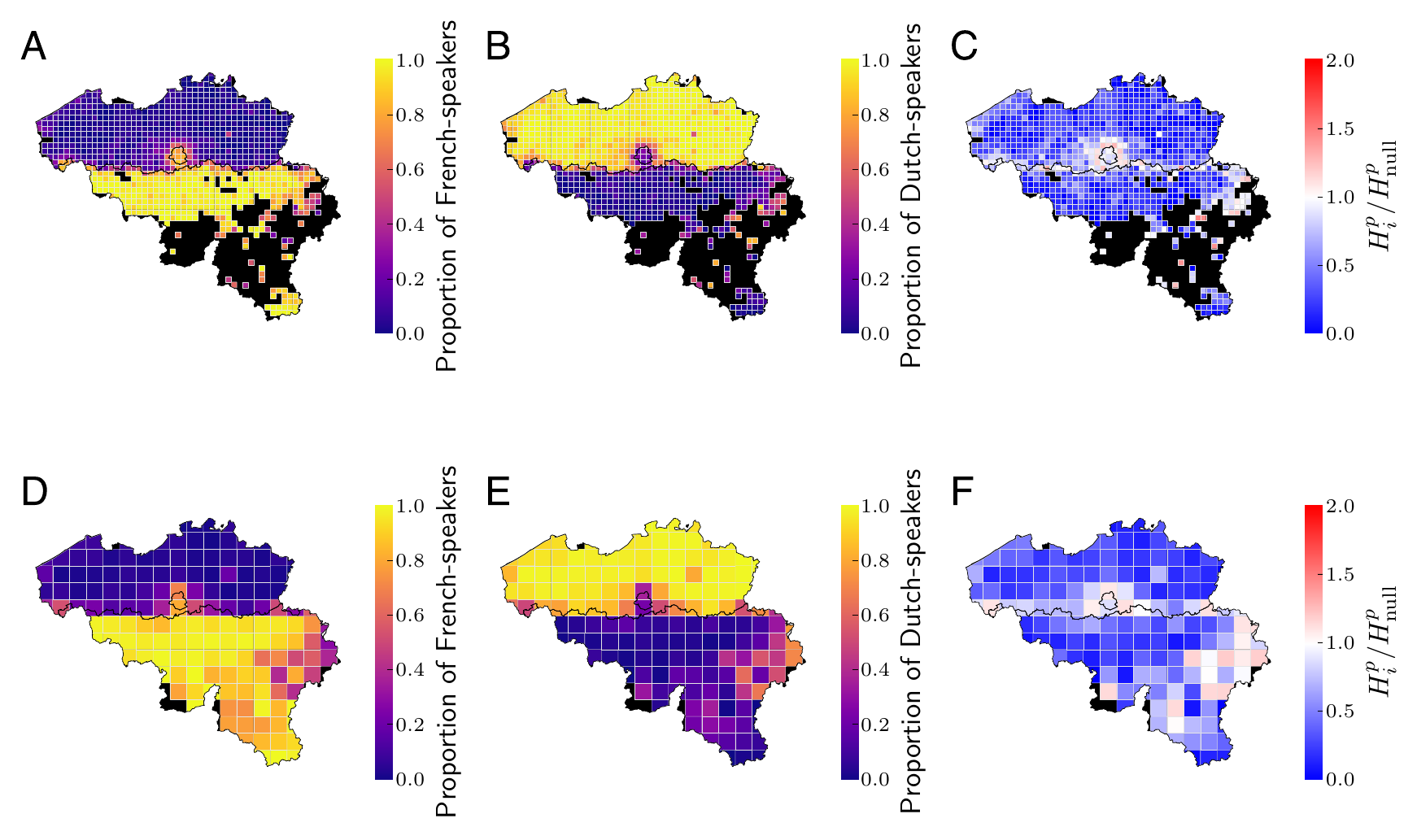}
  \caption{Mapping of the local languages in Belgium from Twitter data. For each cell of $5 \times 5 \, \si{\kilo \meter \squared}$ (\textit{A-B-C}) and $15 \times 15 \, \si{\kilo \meter \squared}$ (\textit{D-E-F}), the proportions $p_{l,i}$ of (\textit{A-D}) French and (\textit{B-E}) Dutch speakers are shown, as well as (\textit{C-F}) the relative proportion entropy of the $L$ groups. The border between Flanders (North) and Wallonia (South) is drawn, and the Brussels Region too. A clear separation of language groups is visible in Belgium following the linguistic regions, displaying mixing mainly around the border and in Brussels. The change of cell size does not radically change the pattern of mixing that we can observe.}
  \label{fig:be_cell_sizes}
\end{figure}

\begin{figure}[p]
  \centering
  \includegraphics{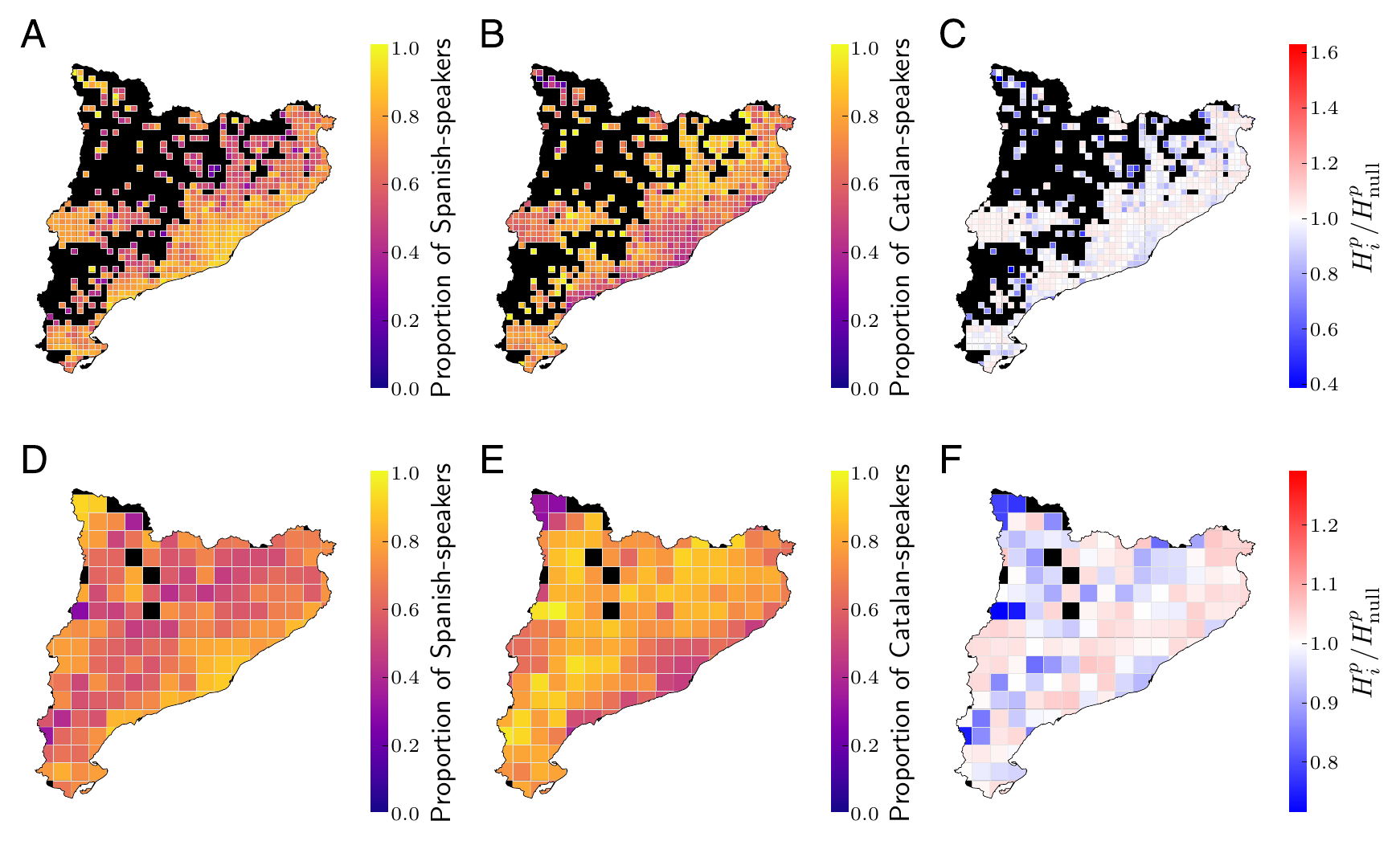}
  \caption{Mapping of the local languages in Catalonia from Twitter data. For each cell of $5 \times 5 \, \si{\kilo \meter \squared}$ (\textit{A-B-C}) and $15 \times 15 \, \si{\kilo \meter \squared}$ (\textit{D-E-F}), the proportions $p_{l,i}$ of (\textit{A-D}) Catalan and (\textit{B-E}) Spanish speakers are shown, as well as (\textit{C-F}) the relative proportion entropy of the $L$ groups. A rather good mixing of languages appears for both cell sizes. The change of cell size does not radically change the pattern of mixing that we can observe.}
  \label{fig:cat_cell_sizes}
\end{figure}

\begin{figure}[p]
  \centering
  \includegraphics{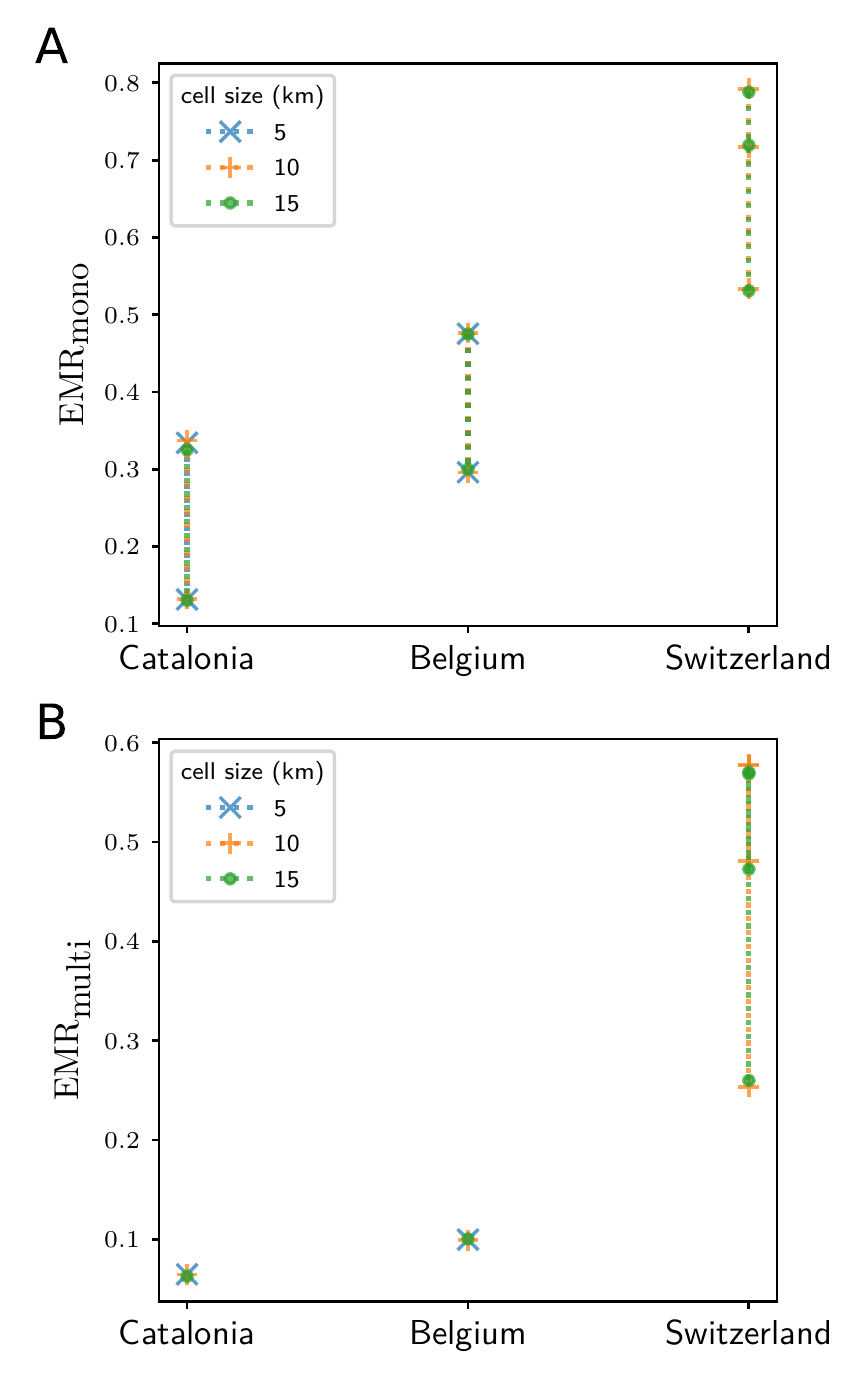}
  \caption{Robustness of the Earth Mover's Ratio with regards to the choice of cell size. For cell sizes ranging from $5 \times 5 \, \si{\kilo \meter \squared}$ to $15 \times 15 \, \si{\kilo \meter \squared}$ in Catalonia, Belgium and Switzerland, the values of the EMR (\textit{A}) between the monolinguals in each language and the whole population, and (\textit{B}) between the multilinguals and the whole population are shown.}
  \label{fig:EMR_cell_size_invariance}
\end{figure}

Usually, a range of cell sizes is acceptable, and although we chose to show only one in the main text for brevity, our analysis can also be carried out for other sizes. We thus show in Figs. \ref{fig:be_cell_sizes} and \ref{fig:cat_cell_sizes} the maps of languages' proportions and relative entropy in Belgium and Catalonia for cells of $5 \times 5 \, \si{\kilo \meter \squared}$ and $15 \times 15 \, \si{\kilo \meter \squared}$. We further show how our measure of segregation through the Earth Mover's Ratio (EMR) is robust to reasonable cell size changes in Fig. \ref{fig:EMR_cell_size_invariance}.

\subsection{EMR reliability}
\label{sec:emr_reliability}
We give the results of EMR calculations in Table \ref{tab:region_summary}, but seeing the relatively low counts of users found in some groups, one may wonder whether such samples may yield reliable measures of segregation through the EMR. Therefore, to determine whether the measure of the EMR for a group can be deemed reliable, we first set a hard minimum of 50 users detected in that group. Then, we order the cells by descending concentration of the whole population, and take as group count threshold the inverse of the concentration in the cell corresponding to the $90^\textrm{th}$ percentile of the cumulative distribution. This way, the sample we have can be expected to have been sufficient to populate significant cells. However, this threshold may not be passed for a minority group localized in low-density areas, while we may still have a sufficient sample relatively to its actual size. For this reason, we also test whether the EMR calculation is robust to bootstrap resampling, as we generate 50 samples from the concentration distribution of this group, calculate the EMR each time, and if the relative standard deviation of these is below 10\%, we consider the measured EMR of the group to be reliable. We have thus determined that the EMR calculated for the bilinguals in Cyprus, all the multilingual groups in Luxembourg and the trilinguals in Switzerland were not reliable.

\section{Results from the Canadian census}
\label{sec:canada_census}
\begin{figure}[p]
  \centering
  \includegraphics{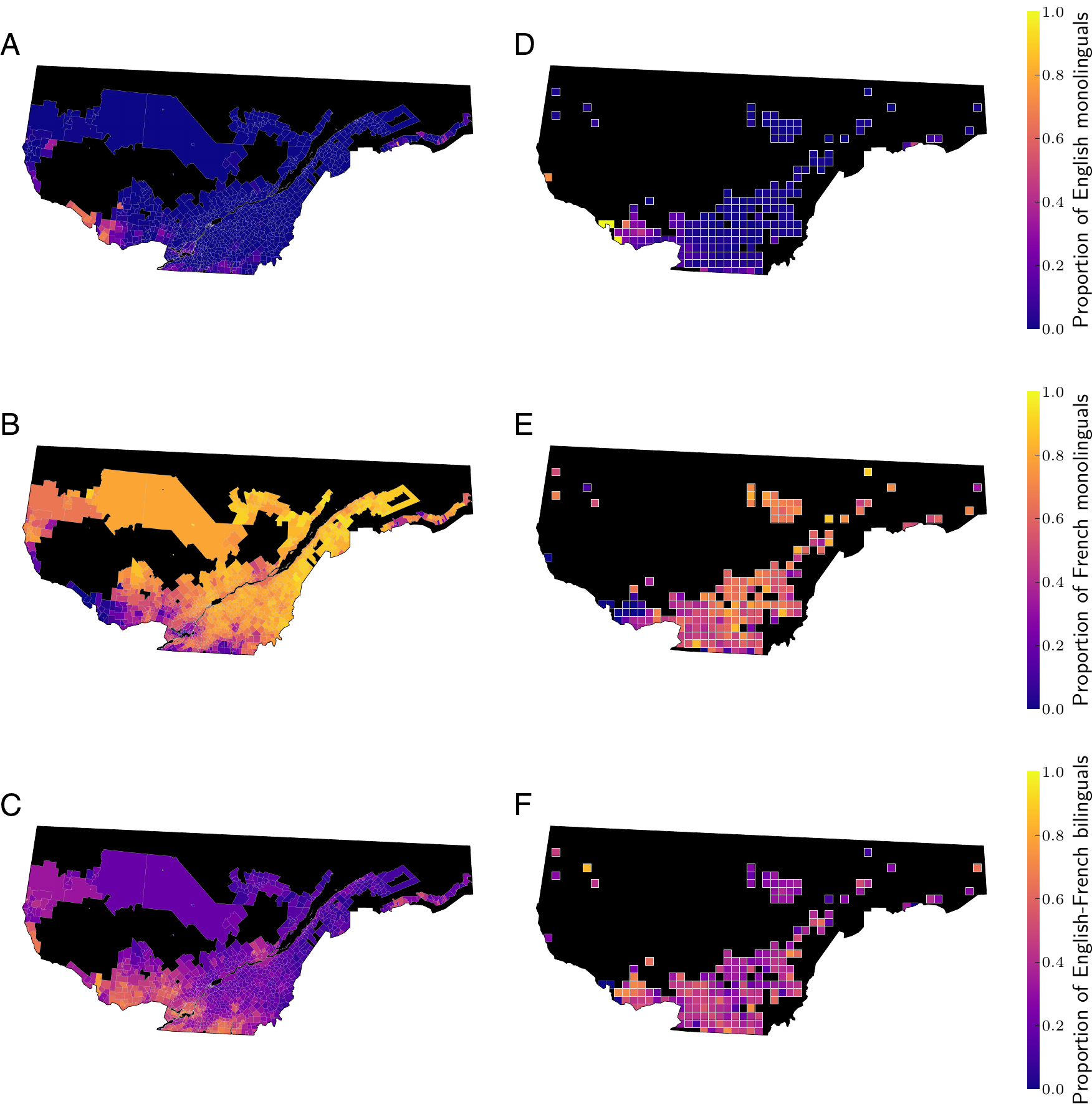}
  \caption{Map of the proportions of $L$-speakers in Quebec. (\textit{A}-\textit{B}-\textit{C}) Drawn using data from the Canadian census of 2016 about the knowledge of the languages, with results on the scale of census subdivisions, for respectively English monolinguals, French monolinguals and bilinguals. (\textit{D}-\textit{E}-\textit{F}) Drawn using Twitter data. In both cases, we can see English monolinguals are only present in relevant numbers near Ottawa (South West), that bilinguals are predominant in the cities of Montreal and Quebec, and French monolinguals are in the countryside.}
  \label{fig:comp_prop_qbc}
\end{figure}

\begin{figure}[p]
  \centering
  \includegraphics{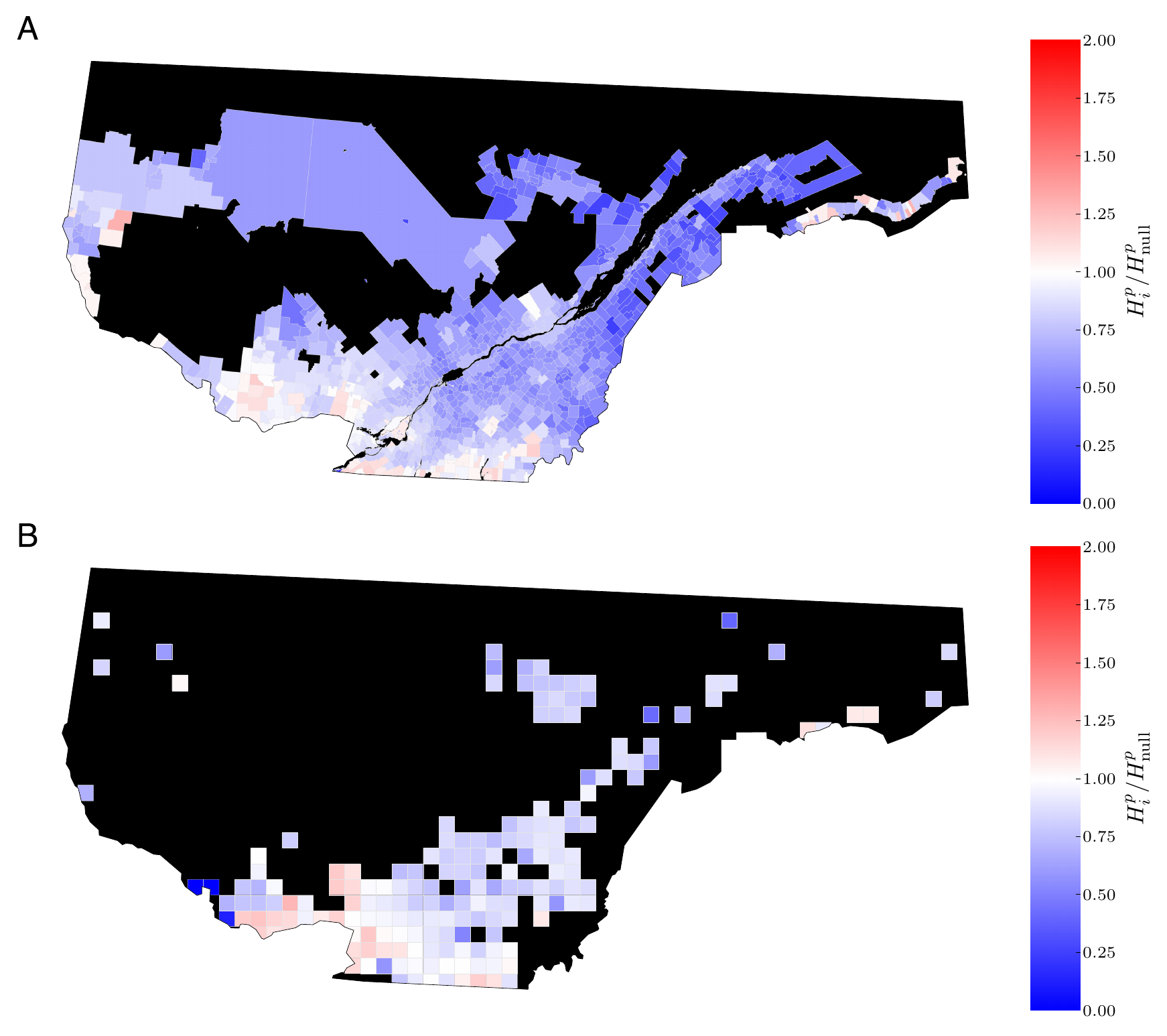}
  \caption{Map of the relative proportion entropy of the $L$ groups in Quebec. (\textit{A}) Drawn using data from the Canadian census of 2016 about the knowledge of the languages, with results on the scale of census subdivisions. (\textit{B}) Drawn using Twitter data. Mixing is greater in the largest cities and near the southern border, while there is more segregation close to Ottawa (South West) and in the countryside.}
  \label{fig:comp_Hp_qbc}
\end{figure}

We have obtained from the 2016 Canadian census data which enabled us to map languages on a fine enough spatial scale. Equivalent data could not be found for other countries. For instance, in Belgium the census cannot contain questions about language by law \cite{VuyeLanguageTerritoriality2010}. Other censuses either only provide global numbers, or ones not considering bilingualism. In Quebec, the question asked was ``Can this person speak English or French well enough to conduct a conversation?'', and the answers that could be selected were:
\begin{itemize}
  \item English only
  \item French only
  \item Both English and French
  \item Neither English nor French
\end{itemize}
(see the full questionnaire at \href{https://www12.statcan.gc.ca/nhs-enm/2016/ref/questionnaires/questions-eng.cfm}{https://www12.statcan.gc.ca/nhs-enm/2016/ref/questionnaires/questions-eng.cfm}). We extracted the census data from the census subdivisions within Quebec and were thus able to get the results shown in Figs. \ref{fig:comp_prop_qbc} and \ref{fig:comp_Hp_qbc}. To simplify the visualization and computations, we chose to crop out the northern part of Quebec, which is scarcely populated. While one can observe differences in proportions in particular places, the patterns observed are similar between the two data sources.

\section{Considerations leading to our model}
\label{sec:model_dev}
Let us start from a more general form of a bilingual model encompassing the two models presented in the main text, without adding too much complexity. The most general form we can think of in terms of a master equation is as follows:
\begin{equation}
  \begin{aligned}
    \dv{p_A}{t} &= \mu p_{AB} f_{AB \rightarrow A} (p_A, p_{AB}) - c (1 - \mu) p_A f_{A \rightarrow AB} (p_B, p_{AB}) \\
    \dv{p_B}{t} &= \mu p_{AB} f_{AB \rightarrow B} (p_B, p_{AB}) - c (1 - \mu) p_B f_{B \rightarrow AB} (p_A, p_{AB}),
  \end{aligned}
\end{equation}
% why we prefer to keep c there and not multiplying the whole equation: this way mu is fixed and has a physical interpretation, we prefer to fix it and tweak c than the other way around.
where, for $X \in \{ A, B \}$, $f_{AB \rightarrow X}$ and $f_{X \rightarrow AB}$ are polynomials modeling the social pressure associated to learning a language or teaching it or not to one's children. They involve the relevant speakers' proportions, and their coefficients are related to concepts such as prestige or the language preference of bilinguals. Let us write the following simple forms for these polynomials:
\begin{equation}
  \begin{aligned}
    f_{AB \rightarrow X} (p_X, p_{AB}) &= s_X (p_X + q_X p_{AB}) \\
    f_{X \rightarrow AB} (p_Y, p_{AB}) &= s_Y (p_Y + q_Y p_{AB}),
  \end{aligned}
\end{equation}
where $Y$ is the monolingual group opposite of $X$, $s_X$ is the prestige of $X$ and $q_X$ is the preference of bilinguals for speaking $X$ in their conversations. For all proportions and other parameters equal, a higher prestige for $X$ will make bilinguals forget $Y$ more easily, and $Y$ monolinguals learn $X$ more easily than $X$ monolinguals learn $Y$. As for a higher $q_X$, it makes $X$ appear more present through bilinguals. In short, with this model we keep the concept of prestige and we additionally model the bilinguals' influence on the social pressure to speak $X$ or $Y$ language in an asymmetric fashion, as they may favor a given language in their interactions.

A new model can then be written with the following master equation:
\begin{equation}
  \begin{aligned}
    \dv{p_A}{t} &= \mu p_{AB} s_A^+ (p_A + q_A^+ p_{AB}) - c (1 - \mu) p_A s_B^- (p_B + q_B^- p_{AB}) \\
    \dv{p_B}{t} &= \mu p_{AB} s_B^+ (p_B + q_B^+ p_{AB}) - c (1 - \mu) p_B s_A^- (p_A + q_A^- p_{AB}).
  \end{aligned}
\end{equation}
The model simplifies to the original Minett-Wang model \cite{Minett2008} when $q_A^\pm = q_B^\pm = 0$ and $s_B^\pm = 1 - s_A^\pm$, and to a slightly modified version of Castell\'{o}'s model of bilinguals \cite{Castello2006} when $q_A^- = q_B^- = 0$ and $q_A^+ = q_B^+ = 1$. In the rest, we will consider the following simplified version:
\begin{equation}
  \label{eq:bi_pref_global}
  \begin{aligned}
    \dv{p_A}{t} &= \mu p_{AB} s (p_A + q p_{AB}) - c (1 - \mu) p_A (1-s) (p_B + (1-q) p_{AB}) \\
    \dv{p_B}{t} &= \mu p_{AB} (1-s) (p_B + (1-q) p_{AB}) - c (1 - \mu) p_B s (p_A + q p_{AB}),
  \end{aligned}
\end{equation}
which amounts to assuming that prestige and bilingual preference have the same effect on the learning and inheriting process ($q_A^+ = q_A^- \equiv q_A$), and that they are defined complementarily between the two languages ($q_A = 1 - q_B \equiv q$).

Using $p_A + p_B + p_{AB} = 1$ we can write these two in terms of $p_A$ and $p_B$:
\begin{equation}
  \begin{aligned}
    \dv{p_A}{t} &= \mu p_{AB} s (p_A (1-q) + q (1 - p_B)) - c (1 - \mu) p_A (1-s) (q p_B + (1-q) (1 - p_A)) \\
    \dv{p_B}{t} &= \mu p_{AB} (1-s) (q p_B + (1-q) (1 - p_A)) - c (1 - \mu) p_B s (p_A (1-q) + q (1 - p_B)).
  \end{aligned}
\end{equation}

\section{Metapopulation}
\label{sec:metapop}

\subsection{Initialization for the simulations}
\label{sec:metapop_init}
To initialize the state of the system for the simulations, since we are mixing commuting data from census with data from Twitter, we first have to up-scale the number of individuals we found from Twitter data to match the number of commuters. Instead of up-scaling uniformly by the ratio between these two numbers, we choose here to alleviate the biases of Twitter through re-scaling factors $k_{L,i}$, which are different between cells and language groups. The first bias to address is how users are more urban than average \cite{Mislove2011}, which leads us to impose the constraint $\sum_L k_{L,i} N_{L,i} = N_i^\text{(census)}$. The second one is how language usage is different on Twitter, because one's audience is wider, potentially more cosmopolitan than offline \cite{Nguyen2015}, and with different demographics \cite{Mislove2011}. Hence we want to also re-scale such that $\sum_i k_{L,i} N_{L,i} = N_L^\text{(census)}$, with $N_L^\text{(census)}$ the global numbers of $L$-speakers, data which are much more widely available than the counts for each census tract. By imposing these two constraints, it is assumed that the two biases are independent: the bias in the choice of language used online is space-independent in a specific multilingual society, and the over-representation of urban people is $L$-group-independent. The re-scaling was performed using Iterative Proportional Fitting (IPF), an algorithm conceptualized by \cite{Deming1940,Fienberg1970}, and which we implemented with \cite{Forthomme2016}. Once done, a synthetic population of $k_{L,i} N_{L, i}$ residents in every cell $i$ for every $L$-group is created.
The population thus obtained then only lacks a work cell. The census data on commuting used to make this attribution are available on a tract-level. Accordingly, they are first to be translated to the cells we define as the bins of the language distributions. To do so, census tracts are intersected with the cells, and the ratios of the intersected areas with the tracts' whole areas are then calculated. Commuters are distributed proportionally to this ratio in residence and work cells. This pre-processing provides an origin-destination matrix $\bm{\sigma}$, giving for each cell $i$ what proportion of individuals commuted to each cell $j$ (including $j=i$). The synthetic population created in each cell $i$ can then be given a work cell $j$, attributed with a probability distribution defined by the set $\{ \sigma_{ij} \}_j$, uniformly across language groups. We assume that the commuting patterns of individuals is $L$-group-independent, which is a rather reasonable assumption when $L$ groups have a similar socioeconomic status, like in our study cases of Belgium and Catalonia.

\subsection{Equivalent equations}
\label{sec:metapop_analytic}
Here we wish to derive a system of equations describing the evolution of the metapopulation under reasonable assumptions. Let us first rewrite \eqref{eq:bi_pref_global}, the global equations of our model, in terms of counts instead of proportions:
\begin{equation}
\label{eq:bipref_cell_eq}
  \begin{aligned} 
    \dv{N_{A, i}}{t} &= \mu s N_{AB, i}
    \frac{
        \sum_{k}  (N_{A, k} + q N_{AB,k})
    }{
        \sum_{k} N_{k}
    }
    - c (1-\mu) (1-s) N_{A, i}
    \frac{
        \sum_{k} (N_{B,k} + (1-q) N_{AB, k})
    }{
        \sum_{k} N_{k}
    }, \\
    \dv{N_{B, i}}{t} &= \mu (1-s) N_{AB, i}
    \frac{
        \sum_{k} (N_{B, k} + (1-q) N_{AB,k})
    }{
        \sum_{k} N_{k}
    }
    - c (1-\mu) s N_{B, i}
    \frac{
        \sum_{k} (N_{A,k} + q N_{AB, k})
    }{
        \sum_{k} N_{k}
    },
  \end{aligned}
\end{equation}
for every cell $i$.
Let us translate these to a metapopulation level, for which the equations hold for the sub-populations of each cell $i$. We will follow what was done in \cite{Sattenspiel1995}, and divide every population $N_{L,i}$ according to their work destination $j$. We thus introduce the notation $N_{L,ij} (t)$ which is the number of $L$-speakers who are residents in $i$ and are at $j$ for work at time $t$. It is such that
\begin{equation}
  N_{L,i} (t) = \sum_j N_{L, ij} (t).
\end{equation}
Then, the equations we want to solve to get the equilibrium points are, for every $i$,
\begin{equation}
\label{eq:fixed_points_eq}
  \begin{aligned} 
    \dv{N_{A, i}}{t} &\equiv \sum_j \dv{N_{A, ij}}{t} = 0 \\
    \dv{N_{B, i}}{t} &\equiv \sum_j \dv{N_{B, ij}}{t} = 0.
  \end{aligned}
\end{equation}

Regarding commuting, we will here use the notations from \cite{Balcan2010}, and introduce first $\sigma_{ij}$, the commuting rate between the sub-population $i$ and every other cell $j$. The return rate of commuting individuals, that is the inverse of the timescale of their stay at work, is denoted $\tau$. The sub-population size evolution (summing over all languages) due to commuting is then given by
\begin{equation}
\label{eq:subpop_commut_evo}
  \begin{aligned}
    \dv{N_{ii}}{t}
      &= \tau \sum_j N_{ij} (t) - \sum_j \sigma_{ij} N_{ii} (t)  \\
    \dv{N_{ij}}{t}
      &= \sigma_{ij} N_{ii} (t) - \tau N_{ij} (t).
  \end{aligned}
\end{equation}
Then, for monolinguals $A$, we can write the following:
\begin{equation}
  \begin{aligned}
    \dv{N_{A,ii}}{t}
      &= \text{A from every destination $j$ returning to their residence $i$} \\
      &- \text{A from $i$ leaving $i$ for work} \\
      &+ \text{AB from $i$ and currently at $i$ turning A} \\
      &- \text{A from $i$ and currently at $i$ turning AB},
  \end{aligned}
\end{equation}
which gives, using both the commuting part from \eqref{eq:subpop_commut_evo} and the language competition part from \eqref{eq:bipref_cell_eq},
\begin{equation}
\label{eq:N_Aii_evo}
  \begin{aligned}
    \dv{N_{A,ii}}{t} 
      &= \tau \sum_j N_{A, ij} \\
      &- \sum_j \sigma_{ij} N_{A,ii} \\
      &+ \mu s (N_{ii} - N_{A, ii} - N_{B, ii}) \left( \frac{\sum_k (N_{A, ki} + q N_{AB, ki})}{\sum_k N_{ki}} \right)\\
      &- c (1-\mu) (1-s) N_{A, ii}  \left( \frac{\sum_k (N_{B, ki} + (1-q) N_{AB, ki})}{\sum_k N_{ki}} \right).
  \end{aligned}
\end{equation}
and similarly for every $j \neq i$:
\begin{equation}
  \begin{aligned}
    \dv{N_{A,ij}}{t}
      &= \text{A arriving at $j$ coming from their residence $i$} \\
      &- \text{A currently at $j$ returning to their residence $i$} \\
      &+ \text{AB from $i$ and currently at $j$ turning A} \\
      &- \text{A from $i$ and currently at $j$ turning AB},
  \end{aligned}
\end{equation}
which gives
\begin{equation}
\label{eq:N_Aij_evo}
  \begin{aligned}
    \dv{N_{A, ij}}{t} 
      &= \sigma_{ij} N_{A, ii} \\
      &- \tau N_{A, ij} \\
      &+ \mu s (N_{ij} - N_{A, ij} - N_{B, ij}) \left( \frac{\sum_k (N_{A, kj} + q N_{AB, kj})}{\sum_k N_{kj}} \right) \\
      &- c (1-\mu) (1-s) N_{A, ij} \left( \frac{\sum_k (N_{B, kj} + (1-q) N_{AB, kj})}{\sum_k N_{kj}} \right).
  \end{aligned}
\end{equation}

Now when we sum \eqref{eq:N_Aij_evo} over $j$ and add \eqref{eq:N_Aii_evo} to try to solve for the system \eqref{eq:fixed_points_eq}, we first see, as in Ref. \cite{Sattenspiel1995}, that the commuting terms simplify.
For the language competition terms, there remains to estimate the $N_{A, ij}$. We will use another result from \cite{Balcan2010}, where they show that under the assumption that $\forall i, \tau \gg \sigma_i$, we can make the following approximation:
\begin{equation}
  \begin{aligned}
    N_{ii} &= \frac{N_i}{1 + \sigma_i / \tau} \\
    N_{ij} &= \frac{N_i \sigma_{ij} / \tau}{1 + \sigma_i / \tau}.
  \end{aligned}
\end{equation}
Let us introduce $\underline{\underline{\nu}}$, a matrix such that $\forall i, \nu_{ii} = 1 / (1 + \sigma_i / \tau)$ and $\forall i, j \text{ such that } i \neq j, \nu_{ij} = \frac{\sigma_{ij} / \tau}{1 + \sigma_i / \tau}$, so we can rewrite
\begin{equation}
  \forall i,j \quad N_{ij} = N_i \nu_{ij}
\end{equation}
These counts, summed over all languages, are then constant. We can also use this approximation for each language, by identification in the equation below:
\begin{equation}
  N_{A, ij} (t) + N_{B, ij} (t) + N_{AB, ij} (t) = (N_{A, i} (t) + N_{B, i} (t) + N_{AB, i} (t)) \nu_{ij}
\end{equation}
We thus obtain the following equivalent equations under our assumptions:
\begin{equation}
  \begin{aligned}
    \dv{N_{A, i}}{t} 
        &= \mu s (N_i - N_{A, i} - N_{B, i})
        \sum_{j} \nu_{ij} \left( 
            q
            + \frac{
            \sum_{k} ((1-q) N_{A, k} - q N_{B,k}) \nu_{kj}
        }{
            \sum_{k} N_{k} \nu_{kj}
        } \right) \\
        &- c (1-\mu) (1-s) N_{A, i}
        \sum_{j} \nu_{ij} \left( 
        1 - q
        + \frac{
            \sum_{k} (q N_{B,k} - (1-q) N_{A, k}) \nu_{kj}
        }{
            \sum_{k} N_{k} \nu_{kj}
        } \right),
    \end{aligned}
\end{equation}
and
\begin{equation}
  \begin{aligned}
  \dv{N_{B, i}}{t} 
      &= \mu (1-s) (N_i - N_{A, i} - N_{B, i}) 
      \sum_{j} \nu_{ij} \left( 
      1 - q
      + \frac{
          \sum_{k} (q N_{B,k} - (1-q) N_{A, k}) \nu_{kj}
      }{
          \sum_{k} N_{k} \nu_{kj}
      } \right) \\
      &- c (1-\mu) s N_{B, i}
      \sum_{j} \nu_{ij} \left( 
      q
      + \frac{
          \sum_{k} ((1-q) N_{A, k} - q N_{B,k}) \nu_{kj}
      }{
          \sum_{k} N_{k} \nu_{kj}
      } \right) .
  \end{aligned}
\end{equation}

\subsection{Simulation results}
\begin{figure}[p]
\centering
  \includegraphics[width=\textwidth]{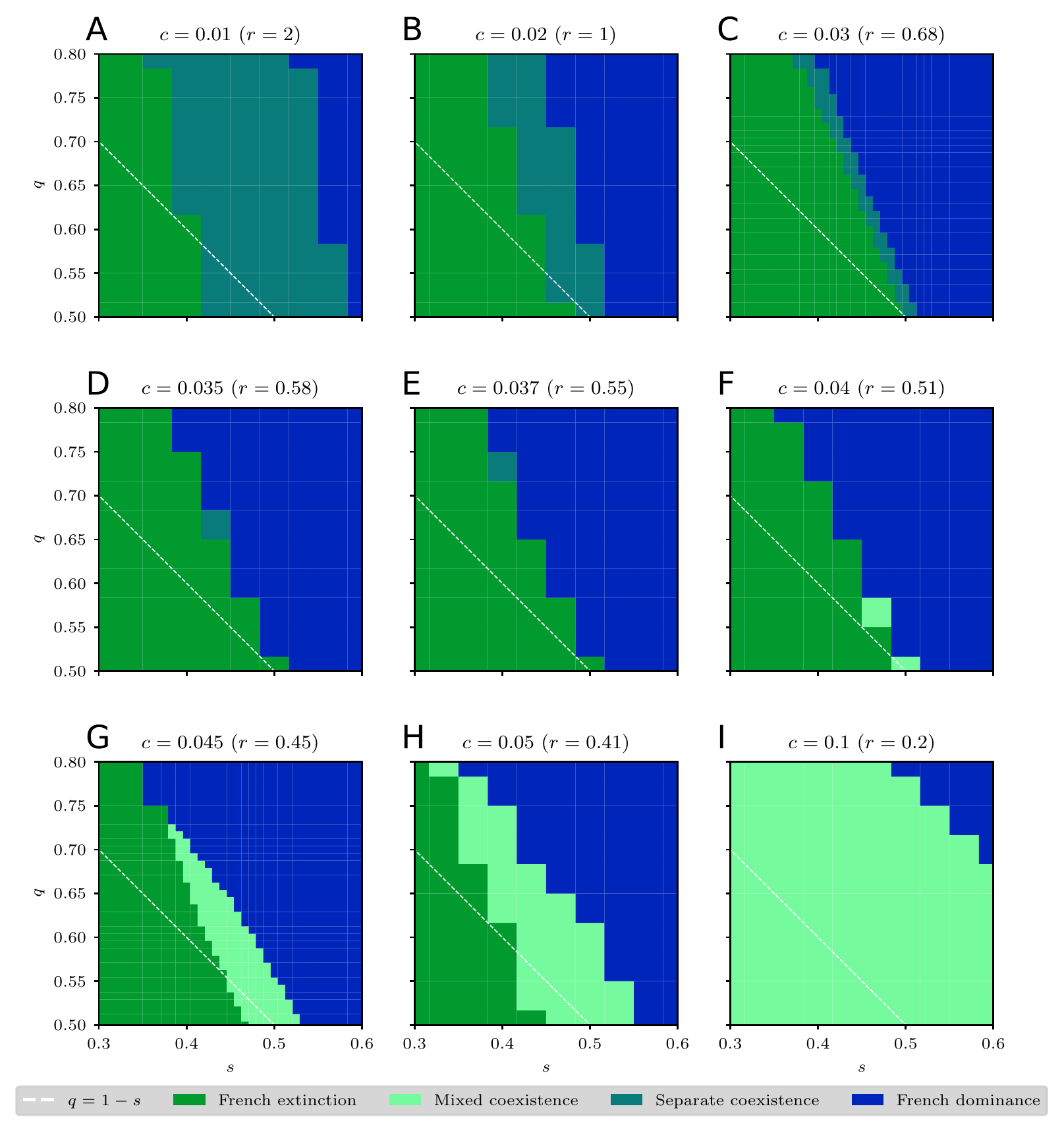}
  \caption{Phase space of the stable states of convergence of our model when iterated within our metapopulation framework in Belgium. Colors show the kind of convergence state reached for the corresponding set of parameters. A lower learning rate, or in other words a lower value of $c$, favors the separate coexistence of French and Flemish, while a higher one favors mixed coexistence. In the transition between the two however, coexistence is almost impossible, as the corresponding region in the $(s,q)$ space becomes very narrow.}
  \label{fig:metapop_phase_space_all_c}
\end{figure}

Starting from the initial population we set up in Belgium, we ran simulations of our model until convergence to a stable state, exploring the parameter space. From these simulations we obtain the regions in the parameter space where each kind of stable state emerge, which are shown in Fig. \ref{fig:metapop_phase_space_all_c}. What is interesting to note here is how a relatively high learning rate favors mixed coexistence while relatively low ones favor separate coexistence, but, surprisingly, in the transition between the two, there are values of $r$ for which all kinds of coexistence are almost impossible.

% Each figure should be on its own page
\begin{table}[p!]
  \centering
  \caption{Number of Twitter users found to be residents and speaking a local language for several regions of interest.}
  \label{tab:region_counts}
  \begin{filecontents*}{data/count_locals.csv}
regionname,countlocals
Balearic islands,13731
Basque country,22120
Belgium,41214
Catalonia,101688
Cyprus,4227
Estonia,2667
Finland,15789
Galicia,30850
Java,840223
Latvia,15502
Luxembourg,1656
Malaysia,347328
Paraguay,45745
Quebec,16848
Switzerland,18552
Valencian Community,59475
\end{filecontents*}
\pgfplotstabletypeset[
    column type=l,
    assign column name/.style={
      /pgfplots/table/column name={\textbf{#1}}},
    every head row/.style={
      after row=\hline\hline},
    columns/regionname/.style={
      string type,
      column name=Region},
    columns/countlocals/.style={
      string type,
      column name={\textbf{Number of local Twitter users}},
      column type={S}},
  ]
  {data/count_locals.csv}
\end{table}
\clearpage

\begin{table}[p]
  \centering
  \caption{Summary metrics for 16 regions of interest. For each language group, either regrouping all speakers of a language including monolinguals and multilinguals or considering mutually exclusive groups, we give the number of residents of that region found to belong to this group based on their tweets. We also give the Earth Mover's Ratio of each group, which measures their spatial segregation in that region.}
  \label{tab:region_summary}
  \begin{filecontents*}{data/summary_table_tex.csv}
Region,Cell size (in km),Group,Count,EMR
Balearic islands,5.0,Spanish-speakers,13046,0.164
,,Catalan-speakers,3839,0.441
,,Catalan monolinguals,685,0.514
,,Spanish monolinguals,9892,0.123
,,Catalan-Spanish bilinguals,3154,0.425
Basque country,5.0,Spanish-speakers,21446,0.025
,,Basque-speakers,4669,0.185
,,Spanish monolinguals,17451,0.063
,,Basque monolinguals,674,0.445
,,Spanish-Basque bilinguals,3995,0.142
Belgium,10.0,French-speakers,14570,0.443
,,Dutch-speakers,27658,0.284
,,French monolinguals,13556,0.476
,,Dutch monolinguals,26644,0.296
,,French-Dutch bilinguals,1014,0.099
Catalonia,10.0,Spanish-speakers,83928,0.045
,,Catalan-speakers,55844,0.151
,,Catalan monolinguals,17760,0.337
,,Spanish monolinguals,45844,0.132
,,Catalan-Spanish bilinguals,38084,0.065
Cyprus,5.0,Greek-speakers,534,0.445
,,Turkish-speakers,3696,0.253
,,Greek monolinguals,531,0.448
,,Turkish monolinguals,3693,0.253
,,Greek-Turkish bilinguals,3,0.607
Estonia,10.0,Estonian-speakers,1987,0.212
,,Russian-speakers,735,0.412
,,Estonian monolinguals,1931,0.226
,,Russian monolinguals,679,0.407
,,Estonian-Russian bilinguals,55,0.506
Finland,40.0,Finnish-speakers,15582,0.044
,,Swedish-speakers,469,0.315
,,Finnish monolinguals,15320,0.042
,,Swedish monolinguals,207,0.31
,,Finnish-Swedish bilinguals,262,0.373
Galicia,10.0,Spanish-speakers,29741,0.008
,,Galician-speakers,13030,0.056
,,Spanish monolinguals,17820,0.045
,,Galician monolinguals,1109,0.151
,,Spanish-Galician bilinguals,11921,0.049
Java,40.0,Indonesian-speakers,837336,0.005
,,Javanese-speakers,74219,0.804
,,Sundanese-speakers,45917,0.373
,,Indonesian monolinguals,726246,0.059
,,Javanese monolinguals,1461,0.686
,,Sundanese monolinguals,1333,0.412
,,Indonesian-Javanese bilinguals,66597,0.833
,,Indonesian-Sundanese bilinguals,38424,0.48
,,Javanese-Sundanese bilinguals,93,0.702
,,Indonesian-Javanese-Sundanese trilinguals,6067,0.515
Latvia,10.0,Latvian-speakers,13431,0.07
,,Russian-speakers,2508,0.304
,,Latvian monolinguals,12994,0.077
,,Russian monolinguals,2071,0.318
,,Latvian-Russian bilinguals,437,0.303
Luxembourg,5.0,French-speakers,1381,0.156
,,German-speakers,292,0.348
,,Luxembourgish-speakers,183,0.22
,,German monolinguals,178,0.411
,,French monolinguals,1262,0.183
,,Luxembourgish monolinguals,56,0.276
,,German-French bilinguals,31,0.214
,,German-Luxembourgish bilinguals,41,0.466
,,French-Luxembourgish bilinguals,44,0.163
,,German-French-Luxembourgish trilinguals,41,0.351
Malaysia,20.0,Malay-speakers,302397,0.037
,,English-speakers,304723,0.019
,,Chinese-speakers,8254,0.208
,,English monolinguals,37919,0.257
,,Malay monolinguals,41302,0.159
,,Chinese monolinguals,1225,0.248
,,English-Malay bilinguals,259853,0.019
,,English-Chinese bilinguals,5786,0.203
,,Malay-Chinese bilinguals,77,0.357
,,English-Malay-Chinese trilinguals,1165,0.198
Paraguay,40.0,Spanish-speakers,45635,0.042
,,Guarani-speakers,3976,0.114
,,Spanish monolinguals,41768,0.04
,,Guarani monolinguals,110,0.558
,,Spanish-Guarani bilinguals,3866,0.116
Quebec,20.0,English-speakers,12534,0.105
,,French-speakers,9971,0.191
,,English monolinguals,6877,0.278
,,French monolinguals,4314,0.304
,,English-French bilinguals,5657,0.113
Switzerland,10.0,German-speakers,9145,0.514
,,French-speakers,8096,0.685
,,Italian-speakers,1930,0.699
,,German monolinguals,8697,0.534
,,French monolinguals,7637,0.717
,,Italian monolinguals,1628,0.792
,,German-French bilinguals,287,0.253
,,German-Italian bilinguals,131,0.578
,,French-Italian bilinguals,142,0.577
,,German-French-Italian trilinguals,28,0.481
Valencian Community,10.0,Spanish-speakers,58413,0.013
,,Catalan-speakers,8567,0.356
,,Catalan monolinguals,1061,0.454
,,Spanish monolinguals,50908,0.048
,,Catalan-Spanish bilinguals,7504,0.342
\end{filecontents*}
\pgfplotstabletypeset[
    skip rows between index={55}{\pgfplotstablerows},
    column type=r,
    assign column name/.style={/pgfplots/table/column name={\textbf{#1}}},
    every head row/.style={after row=\hline\hline},
    every nth row={10}{before row=\hline},
    every row no 5/.style={before row=\hline},
    every row no 15/.style={before row=\hline},
    every row no 25/.style={before row=\hline},
    every row no 35/.style={before row=\hline},
    columns/Region/.style={string type},
    columns/{Cell size (in km)}/.style={int detect},
    columns/Group/.style={string type},
    columns/Count/.style={int detect},
    columns/EMR/.style={fixed, fixed zerofill, precision=3, column type=r},
  ]
  {data/summary_table_tex.csv}
\end{table}

\begin{table}[p]
  \centering
  \pgfplotstabletypeset[
    skip rows between index={0}{55},
    column type=r,
    assign column name/.style={/pgfplots/table/column name={\textbf{#1}}},
    every head row/.style={after row=\hline\hline},
    every nth row={10}{before row=\hline},
    every row no 25/.style={before row=\hline},
    columns/Region/.style={string type},
    columns/{Cell size (in km)}/.style={int detect},
    columns/Group/.style={string type},
    columns/Count/.style={int detect},
    columns/EMR/.style={fixed, fixed zerofill, precision=3, column type=r},
  ]
  {data/summary_table_tex.csv}
\end{table}

\clearpage

\begin{figure}[p]
  \centering
  \includegraphics{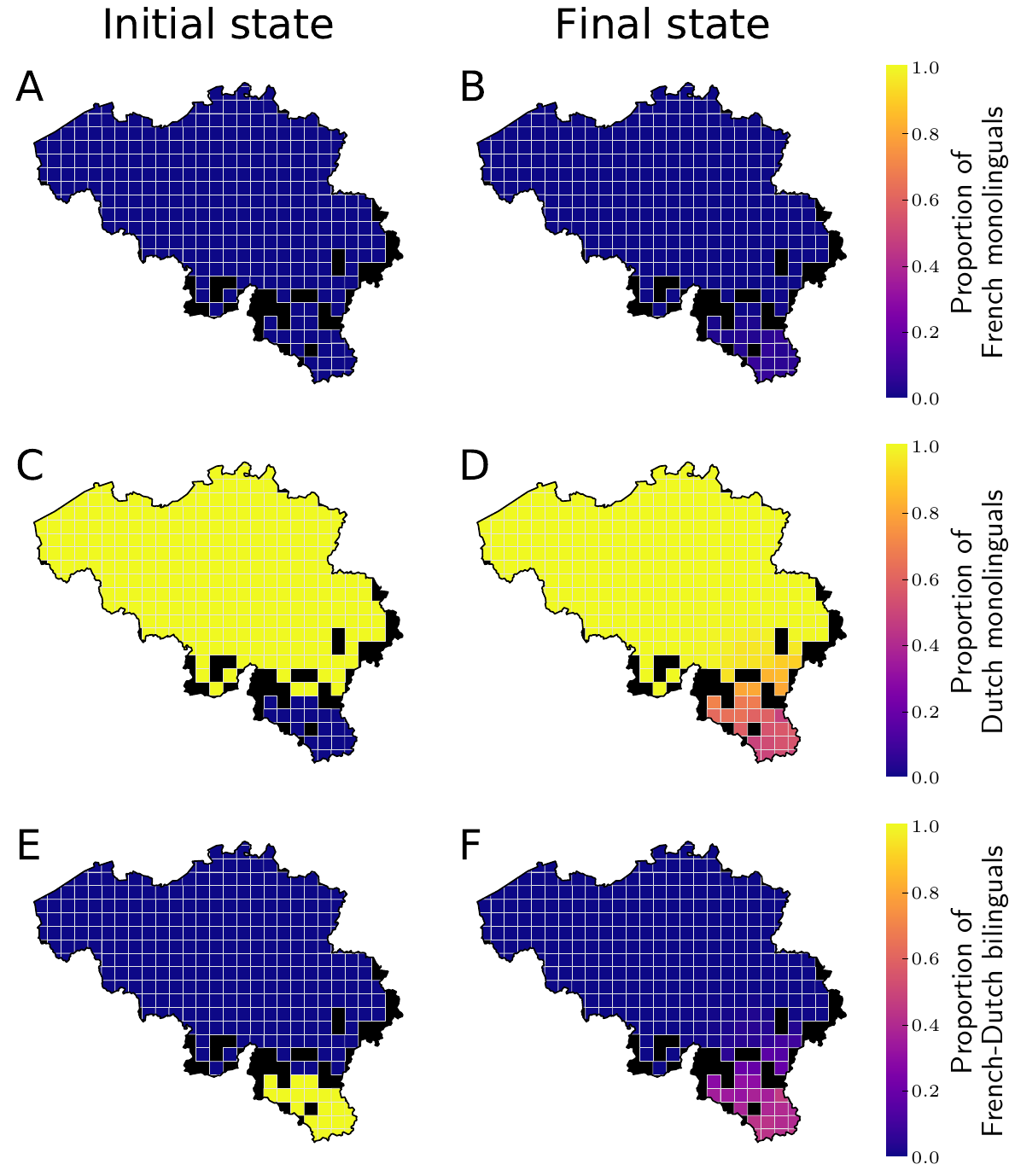}
  \caption{Results of the metapopulation simulation in Belgium with $q$ varying in space. We fixed $s = 0.43$, $c = 0.05$ and $\mu = 0.02$. The system is initialized with no French monolinguals (\textit{A}), only Dutch monolinguals in most of the country, for whom $q = 0.5$ (\textit{C}), except in a small area in the South populated by bilinguals for whom $q = 0.62$ (\textit{E}). The stable state of convergence reached by the simulation features close to no French monolinguals, and only Dutch monolinguals except around the pocket shown in (\textit{E}) where they mix with bilinguals, as shown in (\textit{B}), (\textit{D}) and (\textit{F}), respectively. This demonstrates the possibility for the model to produce a coexistence region within an otherwise monolingual region.}
  \label{fig:spatial_q}
\end{figure}

%%% Add this line AFTER all your figures and tables
\FloatBarrier

\subsection*{Movie S1}
We start from the system initialized with the distributions obtained from Twitter data, and iterate our model with $c=0.005$ and $s= q=1/2$. It first converges to a stable state with a boundary. After 2200 steps, we then increase $c$ by $0.005$ every $400$ steps until we reach $c=0.055$. The system converges quickly to a state of mixed coexistence, with a majority of bilinguals and equal proportions of monolinguals. $c$ is then decreased at the same rate as before to reach its initial value of $0.005$. The system eventually reaches a state displaying a boundary, but displaced compared to the initial boundary.
\bibliography{biblio}